\documentclass[
superscriptaddress,
preprint,
prd,tightenlines,showpacs,nofootinbib,
eqsecnum,
amsfonts,amsmath,amssymb]{revtex4-1}

\usepackage{bm}
\usepackage{graphicx} 
\usepackage{hyperref}
\usepackage{mathrsfs}
\usepackage{multirow}

\newcommand{\ov}[1]{\overline{#1}}
\newcommand{\undl}[1]{\underline{#1}}

\newcommand{\ud}{\mathrm{d}}
\newcommand{\uD}{\mathrm{D}}

\newcommand{\calO}{\mathcal{O}}
\newcommand{\calF}{\mathcal{F}}

\newcommand{\ph}[1]{\phantom{#1}}
\newcommand{\vph}[1]{\vphantom{#1}}

\newcommand{\be}{\begin{equation}}
\newcommand{\ee}{\end{equation}}

\newcommand{\exch}{1\leftrightarrow 2}
\newcommand{\nn}{\nonumber}

\def\beq{\begin{equation}}
\def\eeq{\end{equation}}
\def\bea{\begin{eqnarray}}
\def\eea{\end{eqnarray}}

\def\reff@jnl#1{{\rm#1\/}}
\def\prd{\reff@jnl{Phys. Rev. D }}        
\def\cqg{\reff@jnl{Class. Quantum Grav. }} 

\def\a{\alpha}
\def\b{\beta}

\def\e{\epsilon}
\def\h{\eta}
\def\g{\gamma}
\def\l{\lambda}
\def\k{\kappa}
\def\m{\mu}
\def\n{\nu}

\def\r{\rho}
\def\s{\sigma}
\def\t{\tau}

\def\pA{p^{\a}}
\def\pa{p_{\a}}
\def\pB{p^{\b}}

\def\pn{p_{\n}}
\def\uA{u^{\a}}
\def\ua{u_{\a}}
\def\uB{u^{\b}}
\def\ub{u_{\b}}
\def\ug{u_{\g}}
\def\um{u_{\m}}
\def\uL{u^{\l}}
\def\uM{u^{\m}}

\def\ubrb{u^{[\b}}

\def\fct{\frac{4c}{3}}
\def\tm{\tilde{m}}

\def\SMN{S^{\m\n}}
\def\SAB{S^{\a\b}}
\def\Sab{S_{\a\b}}
\def\SAG{S^{\a]\g}}

\def\SM{S^{\m}}

\def\RAlmn{R^{\a}_{\;\;\l\m\n}}

\def\Rblmn{R_{\b\l\m\n}}
\def\RGlmn{R^{\g}_{\;\;\l\m\n}}
\def\Rglmn{R_{\g\l\m\n}}
\def\Rasm{R^{[\a}_{\;\;\;\l\m\n}}

\def\Raa{R^{\a]}_{\;\;\;\l\m\n}}

\def\JRLMN{J^{\r\l\m\n}}
\def\JGLMN{J^{\g\l\m\n}}
\def\Jasm{J^{\b]\l\m\n}}

\def\Jaa{J^{\a]\l\m\n}}

\def\gr{\nabla_{\r}}

\def\OS{{\mathcal O}(S)}
\def\OSS{{\mathcal O}(S^2)}
\def\OSSS{{\mathcal O}(S^3)}
\def\OSSSS{{\mathcal O}(S^4)}

\def\lcs{\e_{\a\b\m\n}}
\def\lct{\h_{\a\b\m\n}}
\def\ilcs{\e^{\a\b\m\n}}

\usepackage{ulem}
\usepackage[usenames]{color}

\allowdisplaybreaks

\begin{document}

\title{Quadratic-in-spin effects in the orbital dynamics and gravitational-wave
energy flux of compact binaries at the 3PN order}

\author{Alejandro \textsc{Boh\'e}}\email{alejandro.bohe@aei.mpg.de}
\affiliation{Albert Einstein Institut,\\
Am Muehlenberg 1, 14476 Potsdam-Golm, Germany}

\author{Guillaume \textsc{Faye}}\email{faye@iap.fr}
\affiliation{Institut d'Astrophysique de Paris,\\ UMR 7095 du CNRS,
  Universit\'e Pierre \& Marie Curie,\\
  98\textsuperscript{bis} boulevard Arago, 75014 Paris, France}

\author{Sylvain \textsc{Marsat}}\email{smarsat@umd.edu}
\affiliation{Maryland Center for Fundamental Physics \& Joint
  Space-Science Center,\\
Department of Physics, University of Maryland, College Park, MD 20742, USA}
\affiliation{Gravitational Astrophysics Laboratory, NASA Goddard Space Flight 
Center, Greenbelt, MD 20771}

\author{Edward K. \textsc{Porter}}\email{porter@apc.univ-paris7.fr}
\affiliation{Fran\c{c}ois Arago Centre, APC, Universit\'e Paris Diderot,\\
CNRS/IN2P3, CEA/Irfu, Observatoire de Paris, Sorbonne Paris Cit\'e,\\
10 rue A. Domon et L. Duquet, 75205 Paris Cedex 13, France}

\date{\today}

\begin{abstract}
  We investigate the dynamics of spinning binaries of compact objects at the
  next-to-leading order in the quadratic-in-spin effects, which corresponds to
  the third post-Newtonian order (3PN). Using a Dixon-type multipolar
  formalism for spinning point particles endowed with spin-induced quadrupoles
  and computing iteratively in harmonic coordinates the relevant pieces of the
  PN metric within the near zone, we derive the post-Newtonian equations of
  motion as well as the equations of spin precession. We find full equivalence
  with available results. We then focus on the far-zone field produced by
  those systems and obtain the previously unknown 3PN spin contributions to
  the gravitational-wave energy flux by means of the multipolar
  post-Minkowskian (MPM) wave generation formalism. Our results are presented
  in the center-of-mass frame for generic orbits, before being further
  specialized to the case of spin-aligned, circular orbits. We derive the
  orbital phase of the binary based on the energy balance equation and briefly
  discuss the relevance of the new terms.
\end{abstract}


\maketitle


\section{Introduction}\label{sec:intro}

Coalescing binary systems composed of stellar mass black-holes and/or neutron
stars are among the most promising sources for a first direct detection of
gravitational waves (GW) by the network of ground-based interferometers formed
by GEO-HF~\cite{geo} and the advanced version of the detectors
LIGO~\cite{ligo} and Virgo~\cite{virgo}, which should resume their science
runs from 2015, approaching gradually their design sensitivity, expected to be
better by an order of magnitude than that of the first generation. The
cryogenic detector KAGRA~\cite{kagra} will join them in a near future. Further
ahead, the space-based observatory eLISA~\cite{Whitepaper, NGOScience} --- a
serious proposal for the mission recently announced by the European Space
Agency --- will allow us to scan a different frequency band where we expect to
detect, notably, GW emitted by supermassive black-hole binaries before merger.

Extraction of the signal from the noisy data by means of matched filtering
techniques and source parameter estimation both require an accurate modeling
of the waveform. For binary systems of compact objects, the inspiralling phase
of the coalescence can be modeled extremely well by resorting to the
perturbative post-Newtonian (PN) scheme (see~\cite{Bliving} for a review), in
which all quantities of interest are expanded as formal series in powers of
$1/c$. For non-spinning (NS) systems, the phase of the waveform is currently
known up to the order 3.5PN (i.e.\ including corrections up to $1/c^7$),
whereas the full polarizations have been obtained up to the order
3PN~\cite{BFIS08} (with the dominant quadrupole and octupole modes in the
decomposition of the waveform in spin-weighted spherical harmonics known up to
the order 3.5PN~\cite{FMBI12,FBI14}).

In recent years, motivated by astrophysical observations suggesting that black
holes in our universe can have significant spins, considerable effort has been
devoted to investigating higher order corrections to the spin effects in the
binary dynamics, mostly restricted to the conservative piece of the body
evolution in the near zone. While for the neutron stars observed so far, the
largest dimensionless spin magnitude ever measured~\cite{Hessels+06} is only
$\chi \sim 0.4$ (and may reasonably be assumed to be much smaller for
typical expected observations), the spin of a black-hole might be commonly
close to its maximal value~\cite{Gou+11,Nowak+12,Brenneman+11,Reynolds13}.
Then, its effect on the waveform can be fairly strong and, in particular, for
spins misaligned with the orbital angular momentum of the system, the dynamics
becomes much more involved as the orbital plane undergoes precession,
resulting in large modulations of the waveforms~\cite{K95,ACST94}. Even in the
simpler case where the spins are aligned with the orbital angular momentum,
they significantly affect the inspiral rate of the binary, i.e.\ the frequency
evolution of the signal, starting at the 1.5 PN order (see for instance
Ref.~\cite{Nitz+13} for a detailed study of the effect of the spin on the
waveform quantified in terms of figures of merit relevant to data analysis).
To make all factors $1/c$ appear explicitly in this paper, we rescale the
physical spin variable $S_{\rm physical}$ as
\be S=c\, S_{\rm physical}=Gm^2\chi \,,
\ee 
where $\chi$ is the dimensionless spin, with value 1 for an extremal Kerr
black hole.

The calculation of the spin PN corrections to the conservative part of the
dynamics and, to some extent, to the radiation field of the binary beyond the
leading order contributions has been tackled using essentially three different
approaches: (i) a Hamiltonian approach that strongly relies on the use of the
(second) Arnowitt-Deser-Missner (ADM) gauge~\cite{SS09a}, and in which the
dissipative part of the dynamics, demanding a special treatment, is generally
discarded (see however Ref.~\cite{WSZS11}), (ii) an effective field theory
(EFT) Lagrangian formalism~\cite{Porto06,PR08a}, whose application to binary
systems in general relativity has been actively developed since the
mid-2000's, and (iii) a post-Newtonian iteration scheme in harmonic
coordinates (PNISH), reviewed in Ref.~\cite{Bliving}, which we follow in the
present paper. The existence of those three independent methods permits
important checks of calculations that are often tedious, whenever quantities
are available at the same order in more than one formalism.

The binary dynamics at the spin-orbit level (i.e.\ linear-in-spin effects,
which will be referred to as SO from now on) are known up to the order 3.5PN
in both the PNISH and ADM
approaches~\cite{FBB06,MBFB13,BMFB13,DJSspin,HS11so,HSS13}, and to the order
2.5PN in the EFT framework~\cite{Porto10,Levi10}. On the other hand,
quadratic-in-spin corrections (labeled as SS throughout the paper) have been
obtained to the order 2PN in the PNISH formalism~\cite{KWW93,Poisson97,BFH12},
while in both the ADM and EFT formalisms they are known up to the order
3PN~\cite{HS11ss,PR08b,HSS12,LS14a}, and even 4PN for the simpler $S_1 S_2$
interactions~\cite{HS11ss,PR08a,Levi08,Levi12}. Higher-order-in-spin
corrections have also been recently derived~\cite{HS07,HS08,Marsat14,LS14a}.
As for the spin contributions to the radiation field, they have mostly been
computed by using the same usual combination of the MPM and PNISH approaches as
in the present paper, although partial results required for the calculation of
the 3PN flux~\cite{PRR10} and the 2.5PN waveform~\cite{PRR12} have been
obtained within the EFT approach. The energy flux of gravitational-wave
radiation is known up to the order 4PN at the SO
level~\cite{BBF06,BMB13,MBBB13}, whereas at the SS level only the leading
order (2PN) terms were known until now~\cite{BFH12}. Moreover, the leading
order cubic-in-spin terms, which arise at 3.5PN, have been calculated very
recently~\cite{Marsat14}.

Our goal here will be to determine, within the PNISH approach, the 3PN
(i.e.\ next-to-leading order) spin-spin corrections entering both the source
dynamics (thereby providing an additional confirmation of the ADM and EFT
results already available at this order) and most importantly the energy flux,
thus completing the knowledge of all the spinning corrections to the phasing
formula up to the 3PN order. At the next order 3.5PN, the only remaining
unknown terms all come from a SS tail contribution. By contrast, the spin
corrections to the full gravitational-wave polarizations are only known to the
poorer 2PN accuracy~\cite{ABFO09,BFH12} and we postpone to future work the
task of obtaining all the corrections up to the order 3PN.

Our source modeling, as well as the one used in the EFT and ADM approaches,
consists in representing each compact object as a (spinning) point particle
whose internal structure is entirely parametrized by a set of effective
multipole moments. The validity of this description, which makes the
calculations tractable analytically, relies on (i) the compact character of
the bodies, and (ii) the weak influence of their internal dynamics to their
``global'' motion in general relativity, often referred to as the effacement
principle~\cite{Damour83}. The foundations of this formalism were laid down in
the seminal works of Mathisson~\cite{Mathisson37,
  Mathisson40,Mathisson37repub}. Later Papapetrou~\cite{Papapetrou51spin},
found a particularly simple form for the evolution equations (which comprise
both the equations of motion and of spin precession) for dipolar particles,
i.e.\ at linear order in spins. His derivation was improved and rephrased in
the language of distribution theory by Tulczyjew~\cite{Tulczyjew59}, whose
method --- systematically extensible beyond the dipolar model --- has been
recently applied at the quadrupolar level~\cite{SP10}. The dynamics of point
particles with finite-size effects described by higher multipoles was
thoroughly investigated by Dixon~\cite{Dixon64,Dixon73,Dixon74,Dixon79}, who
constructed an appropriate stress-energy ``skeleton'' to encode information
about the internal structure of the body while, on their side, Bailey \&
Israel proposed an elegant effective Lagrangian formulation~\cite{BI75}.
Recently, Harte~\cite{Harte12} showed how the formalism of Dixon could be
extended to self-gravitating systems, by constructing appropriate effective
momenta and effective multipole moments evolving in some effective metric.

In the present article, we are interested in the quadratic-in-spin
contributions arising from the quadrupolar moment of the compact object in the
case where it is adiabatically induced by the
spin~\cite{Poisson97,PR08b,Steinhoff11,BFH12}, as well as the simpler
contributions coming from products of SO corrections. Because, in our source
model, we replace extended bodies by point particles within a self-gravitating
system, our approach must be regarded as an effective one and supplemented
with some UV regularization procedure. A good choice is known to be
dimensional regularization, with possible need of renormalization. We find
however that, at this order, the so-called pure Hadamard-Schwartz
prescription~\cite{BDE04} is sufficient, i.e.\ that dimensional regularization
is not necessary.

The paper is organized as follows. In Section~\ref{sec:dynamics}, we explain
how the dynamics of a test point particle endowed with a spin-induced
quadrupolar structure moving in a curved background spacetime is described in
the Dixon-Mathisson-Papapetrou formalism. We also write the equations of
evolution for the particle worldline, as well as for the spin, under a
convenient explicit form, and we define a spin vector of conserved Euclidian
norm in terms of which our PN results shall be written. The validity of the
model to describe the body dynamics in self-gravitating binaries is discussed.
In Section~\ref{sec:PNdynamics}, dedicated to the computation of the
next-to-leading order SS contributions to the PN equations of motion, we
present expressions for the conserved energy in the center-of-mass frame, both
for generic orbits and for the restricted case of circular orbits in the
absence of precession. Finally, Section~\ref{sec:PNflux} sketches the
derivation of the next-to-leading order SS contributions to the GW flux and
includes a discussion of the impact of our newly derived terms on the phase
evolution of non-precessing binaries in the frequency band of LIGO and Virgo.
Because of the length of the equations, some results are relegated to
appendices. Appendix~\ref{app:resultseom} gives the explicit expressions for
the relative acceleration and the precession vector in the center-of-mass
frame, and Appendix~\ref{app:moments} shows the relevant SS contributions to
the source moments. We also give the explicit transformation between spin
vector and spin tensor in Appendix~\ref{app:spinvectortensor}, as well as the
correspondence between our results and the ADM ones in Appendix~\ref{app:adm}.

We use the following conventions henceforth: $\calO(n)$ means
$\calO(1/c^{n})$, i.e.\ represents a contribution of the order $(n/2)$PN at
least. Greek indices denote spacetime coordinates, i.e.\ $\mu = 0,1,2,3$,
while Latin indices are used for spatial coordinates, i.e.\ $i=1,2,3$.
Symmetrization and anti-symmetrization are represented by, respectively,
parenthesis and brackets around indices. We adopt the signature $(-,+,+,+)$
and keep explicit both Newton's constant $G$ and the speed of light $c$.
Finally the covariant derivative along the worldline is written as
$\mathrm{D}/(c\,\mathrm{d}\t) = \uM\nabla_{\m}$, where $\uM$ is the four
velocity of the particle, defined such that $\uM\um=-1$.


\section{Dynamics of quadrupolar particles}\label{sec:dynamics}

We shall now introduce the model we have adopted to represent the two spinning
compact objects composing the binary as point particles. In
Section~\ref{subsec:mathpap}, we display the Dixon-Mathisson-Papapetrou
evolution equations for test bodies at quadrupolar order, set the covariant
spin supplementary condition, and discuss its consequences. In
Section~\ref{subsec:eommass}, we rewrite the equations of motion in terms of
the 4-velocity and introduce a conserved mass.
Section~\ref{subsec:defspinvector} presents the construction of a spin vector
with a conserved Euclidean norm and shows the precession equation it
satisfies. Finally, Section~\ref{subsec:selfgrav} explains to what extent the
Dixon-Mathisson-Papapetrou dynamics can be used for the companions of a
self-gravitating binary.


\subsection{The Dixon-Mathisson-Papapetrou framework}
\label{subsec:mathpap}

When describing the dynamics of a binary system of compact objects with masses
$m_A$, $A=1,2$, in the context of the post-Newtonian approximation, it is
physically sound to model the two companions as point particles. Indeed, the
ratio of the radii $R_A\sim G m_A/c^2$ to the body separation $r_{12}$ is of
the order $G m_A/(r_{12}c^2)$, and thus much smaller than 1. The dynamics of
test point-like objects including finite size effects has been investigated
extensively by Dixon~\cite{Dixon64,Dixon73,Dixon74,Dixon79}, who generalized
the Mathisson-Papapetrou equations for spinning
particles~\cite{Mathisson37,Mathisson40,Papapetrou51spin,CP51} by attaching
arbitrary high-order moments to the individual bodies, beyond the monopole and
the current dipole also referred to as the particle spin. It can also be
derived from an effective Lagrangian-type approach for spinning particles,
pioneered by Bailey \& Israel~\cite{BI75} (see also an extensive study for
special relativity in~\cite{HR74}) and later implemented in
EFT~\cite{Porto06,PR08a}, where higher-order moments appear as parametrizing
couplings in the action to the value of the Riemann tensor and its derivatives
on the worldline.

The Dixon-Mathisson-Papapetrou equations of evolution for a spinning particle
with quadrupolar structure read:
\begin{subequations} \label{eq:evolution}
\begin{align}
  \frac{\uD p^{\alpha}}{c\, \ud \tau} & =  
  -\frac{1}{2c}\RAlmn \uL \SMN-\frac{c}{3}\gr\RAlmn
  \JRLMN \label{eq:dpdt} \, ,\\[1ex]
  \frac{\uD S^{\alpha\beta}}{c^2 \ud\tau} & =  2p^{[\a}u^{\b]} +
  \fct \Rasm \Jasm \label{eq:dsdt} \,,
\end{align}
\end{subequations}
where $\pA$ is the 4-momentum of the particle and $\uL = \ud x^{\lambda}/(c\,
\ud \tau)$ the 4-velocity along the world-line. The anti-symmetric spin tensor
$\SMN$ represents the effective 4-angular momentum of the object, while the
(effective) mass and current type quadrupoles are encoded into the Dixon
quadrupolar tensor $\JRLMN$, which is only constrained at this stage to have
the same symmetry properties as $R^{\rho\lambda\mu\nu}$. 

The stress-energy tensor $T^{\alpha\beta}$ of the model can be constructed
after the Tulczyjew procedure, by making the only assumption that its support
is point like with at most two derivatives acting on the Dirac distributions,
in the three following steps~\cite{SP10}: (i) write the most general
symmetric tensor that involves up to two (covariant) derivatives of the
particle scalar density
\be \label{eq:scalar_density}
n = \int_{-\infty}^{+\infty} c\, \ud \tau'
\frac{\delta^{4}(x-y(\tau'))}{\sqrt{-g}} \, ,
\ee
where $\delta^4(x-y(\tau))$ is a 4-dimensional Dirac delta, with $y(\tau)$ the
particle worldline and $x$ the field point; (ii) derive the hierarchy of
equations verified by the coefficients of $n$ in $T^{\m\n}$ due to the
conservation equation $\nabla_\n T^{\m\n}=0$; (iii)
constrain those coefficients by solving all algebraic equations, which leaves
two sets of ordinary differential equations. Identifying these two equations
to Eqs.~\eqref{eq:evolution} yields the expression of $T^{\m\n}$ in terms of
$p^{\mu}$, $S^{\mu\nu}$ and $\JRLMN$:
\begin{align}\label{eq:Tmunuquadrupole}
  T^{\mu\nu} &= n \left[ p^{(\mu}u^{\nu)} c +\frac{1}{3}
    R^{(\mu}_{\ph{\mu}\lambda\rho\sigma}J^{\nu)\lambda\rho\sigma}c^2 \right] \nn \\
  & \quad  - \nabla_{\rho} \left[n\, S^{\rho(\mu}u^{\nu)} \right] - \frac{2}{3} 
  \nabla_{\rho}\nabla_{\sigma}  \left[ n\, c^2 J^{\rho(\mu\nu)\sigma} \right]  \,.
\end{align}
It can be recovered with a smaller amount of calculation, further assuming
that the system dynamics is governed by the effective Lagrangian of Bailey \&
Israel~\cite{BI75}, by differentiating the resulting action with respect to
the metric~\cite{Marsat14}.

As the spin tensor $\SMN$ is anti-symmetric, it actually contains six degrees
of freedom. Moreover, for an isolated body, the space-time components $J^{0i}$
of the total angular momentum $J^{\mu\nu}=\SMN/c$ in an appropriate
asymptotically Minkowkian gauge represent the mass-type dipole of the object,
and can thus always be taken to be zero. Similarly, for a test particle moving
in a gravitational background, three degrees of freedom among those contained
in the effective spin tensor are expected to be non-dynamical. They may be
eliminated by fixing the ``center-of-body'' reference point with the help of
three independent space-time equations, globally referred to as the spin
supplementary condition (SSC). The three remaining degrees of freedom
correspond to the spatial components of the spin vector $\SM$. Various choices
of SSC are possible (see for instance~\cite{KS07}). Here we shall adopt, in
keeping with previous works, the covariant (or Tulczyjew~\cite{Tulczyjew59})
condition
\beq
\SMN\pn = 0\,.
\label{eq:ssc}
\eeq

Assuming that the rotating bodies are always at equilibrium, we can reasonably
expect their moments to depend on their masses, spins, as well as possible
dimensionless parameters that characterize the internal structures. Notably,
the spins may induce mass quadrupoles as they do for Kerr black holes. This
effect produces spin square contributions that must be crucially taken into
account at quadratic order in the spin variables. Tidal fields inside the
bodies may also generate $\ell\ge 2$ multipoles, but their leading order
contribution to the acceleration would be $\sim (R_{1,2}/r_{12})^5= \calO(10)$
for a compact binary, so that they can safely be neglected in the present
work.

As not all degrees of freedom in the Dixon quadrupole are physical, its value
as a function of time cannot be uniquely determined by the internal dynamics
of the body. In the adiabatic approximation, there exists a relation, valid
along the particle worldline, between $\JRLMN$, the 4-velocity $u^\m$ and the
spin tensor $S^{\m\n}$. It can derived from an effective Lagrangian
$L_\mathrm{SS}$ built to be the most general Lagrangian --- modulo
perturbative redefinitions of the gravitational field, terms in the form of a
total time derivative, terms that vanish under some given SSC, and
$\calO(S^3)$ remainders --- with the properties of: (i) being quadratic in
$S^{\mu\nu}$, (ii) depending on $u^\mu$, the metric $g_{\mu\nu}$, (derivatives
of) the Riemann tensor, as well as some parameters characterizing the
object~\cite{Porto06, PR08b, BFH12}. After redefining $p^{\mu}$, $S^{\mu\nu}$,
we find that the stress-energy tensor associated with $L_\mathrm{SS}$
coincides with that of Eq.~\eqref{eq:Tmunuquadrupole} provided $\JRLMN$ is
given by
\beq
\JRLMN = \frac{3\k}{\tm\, c^4}S^{\s[\r}u^{\l]}S_{\s}^{\;[\m}u^{\n]} \,,
\label{eq:Jdef}
\eeq
at any instant. The above expression properly describes the presence of a
non-vanishing spin-induced quadrupole, with the source dependent constant $\k$
representing the quadrupolar polarisability. The mass parameter $\tm$ is
defined by $p^2 \equiv \pa\pA = -\tm^2 c^2$. Notice that $\tilde{m}$ is not a
priori conserved. In fact, as shown below, its time derivative is
quadratic in spin and cannot be consistently ignored at our accuracy
level.

The (contravariant) 4-momentum and 4-velocity of the particle are
proportional when terms beyond linear order in the spins are
neglected: $\pA=\tm \,c\,\uA+\OSS$. Our first step will consist in expressing
$\pA$ as a function of $\uA$ to quadratic order in the spin. We impose that
the derivative along the worldine of the SSC~\eqref{eq:ssc} is zero, insert
the equations of motion~\eqref{eq:dpdt} and~\eqref{eq:dsdt} into the resulting
identity, and use the fact that $\JRLMN\sim\OSS$ whereas $\SAB\ub\sim\OSSS$.
This yields, at quadratic order in spin,
\beq 
\pA = \tm\, c\, \uA - \frac{\SAB\SMN}{2\tm\, c^3}
\uL\Rblmn+\fct\ub\Rasm\Jasm + \OSSS \,.
\label{eq:ps2}
\eeq
We are now in position to write the spin evolution equation in a more explicit
way. In the Lagrangian formalism, the effective linear and angular momenta are
defined in a way that guarantees the conservation of the spin
magnitude~\cite{Steinhoff11,Marsat14}. This conservation law is a remarkable
feature of the spinning-particle dynamics. In our context, it will follow from
Eq.~\eqref{eq:evolution} for some class of supplementary conditions. In fact,
it can indeed be derived explicitly from those equations, for the
form~\eqref{eq:Jdef} of the quadrupole moment and the covariant
SSC~\eqref{eq:ssc}. By substituting the 4-momentum~\eqref{eq:ps2} into
equation~\eqref{eq:dsdt} we get
\bea
\frac{\uD S^{\alpha\beta}}{c^2\ud \tau}& = & 
\fct\left[\Rasm\Jasm + \ug\ubrb\Raa\JGLMN-\ug\ubrb\Jaa\RGlmn\right] 
\nonumber\\ 
& - & \uL\Rglmn\ubrb\frac{\SAG\SMN}{\tm\, c^3} + \OSSS \,.
\eea
If we contract this expression with $\Sab$, we obtain
$\Sab {\mathrm D}\SAB / (c{\mathrm d}\t)\sim\OSSSS$ and, therefore, defining
the spin magnitude as $s^{2}= \Sab\SAB/2$,
\beq
\frac{\ud s}{\mathrm{d}\tau}\sim\OSSS \,.
\eeq
This demonstrates that the spin magnitude is actually conserved at order
$\OSS$.


\subsection{Conserved mass and evolution equations}\label{subsec:eommass}

Our next task is to investigate the issue of mass conservation at quadratic
order in spins. For this purpose, let us compute the time derivative of the mass
parameter $\tm$. Using the equation of motion~\eqref{eq:dpdt} and the Bianchi
identities, we can write
\beq
-\tilde{m}\,c^{2}  \frac{\ud \tilde{m}}{c\, \ud \tau} = 
p_{\alpha} \frac{\uD p^{\alpha}}{c\, \ud \tau} = 
-\frac{\tm\, c^2}{6}\uB\nabla_{\beta}R_{\r\l\m\n}\JRLMN+ \OSSS \,.
\label{eq:dmdt}
\eeq
As the time dependence of $\JRLMN$ is through the 4-velocity and the spin
tensor, i.e.\ $\JRLMN(\t)=\JRLMN(\uB(\t),\SAB(\t))$, the fact that ${\mathrm
  D}\uB/(c\,{\mathrm d}\tau) \sim\OS$ and ${\mathrm D}\SAB/(c\,{\mathrm
  d}\tau)\sim\OSS$ implies the approximate conservation of the Dixon
quadrupole: ${\mathrm D}\JRLMN/(c\,{\mathrm d}\tau)\sim\OSSS$. Now,
substituting $\uB\nabla_{\beta}$ with ${\mathrm D}/(c\, {\mathrm d}\tau)$, we
can write down the equation
\beq
\frac{\ud}{\ud \tau}\left[\tm - \frac{1}{6}R_{\r\l\m\n}\JRLMN\right] = 
\OSSS\,,
\eeq
which finally allows us to define a conserved quantity $m$ as
\beq
 m \equiv \tm - \frac{1}{6}R_{\r\l\m\n}\JRLMN\,. 
\eeq
Hereafter, the constant parameter $m$ will be regarded as the effective mass
of the particle. This mass is the one that appears in all our post-Newtonian
results. By construction, it is conserved, like the spin magnitude.
Substituting the expression~\eqref{eq:dmdt} into Eq.~\eqref{eq:ps2} gives us
the link between the 4-momentum and the 4-velocity:
\be\label{eq:pmutoumu}
\pA = m \,c \, \uA +\frac{c}{6}\uA R_{\r\l\m\n}\JRLMN - 
\frac{\SAB\SMN}{2m\, c^3}\uL\Rblmn + \fct\ub\Rasm\Jasm + \OSSS \,,
\ee
where $m$ was just shown to be a constant parameter at order $\OSS$. We are then in
position to rewrite the evolution equations for spinning particles to
quadratic order in the spins, using the 4-velocity instead of the 4-momentum.
Those are:
\begin{subequations}
\begin{align}
  \frac{\uD\uA}{\ud\t} &= \frac{u^{\rho}}{2}
  \frac{\uD R_{\b\r\m\n}}{\ud\t}\frac{\SAB\SMN}{m^{2}c^{4}} -
  \frac{1}{2}\RAlmn\uL\frac{\SMN}{m\, c}-
  \frac{c}{3}\gr\RAlmn\frac{\JRLMN}{m} \nonumber\\
  & - \frac{\uA}{6}\frac{\uD R_{\l\r\m\n}}{\ud \t}\frac{J^{\l\r\m\n}}{m} - 
  \frac{4\ub}{3} \frac{\uD R^{[\a}_{\;\;\;\r\m\n}}{\ud \t}
  \frac{J^{\b]\r\m\n}}{m}+\OSSS \,,\\[1ex]
  \frac{\uD\SAB}{\ud\t} & = \frac{4c^3}{3}\left[\ubrb
    R^{\alpha]}_{\ph{\alpha}\lambda\mu\nu}\ug\JGLMN - \ubrb\Jaa\ug\RGlmn +
    \Rasm\Jasm  \right] \nonumber\\
  & \qquad - \Rglmn\uL\ubrb\frac{\SAG\SMN}{m\, c} + \OSSS \,.
\end{align}
\end{subequations}
%


\subsection{Definition of a spin vector and equation of precession}
\label{subsec:defspinvector}

From the anti-symmetric spin tensor $\SAB$, we define the spin 4-covector
$\tilde{S}_{\alpha}$ as
\beq\label{eq:defscovector}
\tilde{S}_{\a} = -\frac{1}{2}\lcs\frac{\pB}{m\, c}\SMN \,,
\eeq
where $\lcs = \sqrt{-g}\,\lct$ denotes the covariant Levi-Civita tensor, with
$\lct$ being the completely anti-symmetric symbol that verifies $\eta_{0123} =
1$, and where $g=\det g_{\mu\nu}$ is the determinant of the metric tensor in
generic coordinates. The tilde on this covariant spin vector will allow us to
distinguish it from the Euclidean conserved-norm spin vector we shall
introduce below. Notice that $\tilde{S}_{\a}$ automatically satisfies
$\tilde{S}_{\alpha}\pA=0$ and thus carries 3 degrees of freedom as required.
If we contract the above equation with
$\epsilon^{\alpha\gamma\rho\sigma}p_{\gamma}$ and use the SSC
$p_{\nu}S^{\mu\nu} = 0$, we can invert Eq.~\eqref{eq:defscovector} and obtain
the spin tensor in terms of $\tilde{S}_{\alpha}$:
\beq
\SAB = \ilcs \frac{p_{\m}}{m\,c}\tilde{S}_{\n} + \calO(S^3)\,.
\label{eq:st1}
\eeq
Remembering that
$\pB=m\,c\,\uB+\OSS$ at the linear-in-spin level, it is straightforward to
check that $\tilde{S}_\alpha \tilde{S}^\alpha=s^2$, by virtue of the relation
$\lcs \epsilon^{\lambda\rho\sigma\tau} = -4! 
\delta^{[\lambda}_{\,  \alpha}\delta^{\rho}_{\, \beta}
\delta^{\sigma}_{\, \mu}\delta^{\tau]}_{\, \nu} $.

To derive the evolution equation for the spin 4-covector, we differentiate
Eq.~\eqref{eq:defscovector} with respect to the proper time, which yields
\beq
\frac{\uD \tilde{S}_{\alpha}}{\ud \tau} = 
\lcs\left[\frac{1}{4} R^{\b}_{\;\;\l\s\r}\uL \frac{S^{\s\r}\SMN}{m\, c} - 
\frac{2c^3}{3}R^{\m}_{\;\;\l\s\r}J^{\n\l\s\r}\uB\right] + 
\OSSS \,.
\label{eq:sp1}
\eeq
In what follows, we shall explicitly resort to our particular
form~\eqref{eq:Jdef} for $J^{\m\n\a\b}$, relevant in the case of a
spin-induced quadrupole. It will be convenient to investigate each term on the
right-hand side of Eq.~\eqref{eq:sp1} individually. With our definition of
$\tilde{S}_{\m}$, the first term there reads
\beq
\frac{1}{4} \lcs R^{\b}_{\;\;\l\s\r}\uL \frac{S^{\s\r}\SMN}{m\, c} = 
- \frac{1}{2}u_{\k}\e^{\s\r\k\g}\frac{\tilde{S}_{\g}}{m\, c} 
R^{\b}_{\;\;\l\s\r}\tilde{S}_{\b}\uL u_{\a} + \OSSSS \,.
\eeq
At this stage, it is useful to introduce the gravitomagnetic part of the Bel
decomposition of the Riemann tensor
\beq
H_{\m\n} = 2\,{\vph{R}}^{*}\!R_{\mu\kappa\nu\lambda}u^{\kappa}u^{\lambda} \,.
\eeq
where ${\vph{R}}^{*}\!R$ is the self-dual Riemann tensor defined by
\beq
{\vph{R}}^{*}\!R_{\a\b\m\n} = \frac{1}{2}\e_{\a\b}^{\;\;\;\;\k\g}\,R_{\k\g\m\n} 
\,.
\eeq
Physically, the tensor $H_{\m\n}$ represents the tidal current-type
quadrupole in the relativistic theory of tides.
We can now put Eq.~\eqref{eq:sp1} in the form
\beq
\frac{1}{4}\lcs R^{\b}_{\;\;\l\s\r}\uL \frac{S^{\s\r}\SMN}{m\, c} = 
\frac{1}{2}H^{\g\b}\ua \frac{\tilde{S}_{\g}\tilde{S}_{\b}}{m\, c} + \OSSS \,.
\eeq
Let us focus next on the second expression on the right-hand side of
Eq.~\eqref{eq:sp1}. After substituting the value for $J^{\n\l\s\r}$ therein,
we rewrite the resulting expression in terms of the gravitoelectric part of
the Bel decomposition of the Riemann tensor
\beq
G_{\m\n} = -R_{\m\l\n\r}\uL u^{\r} \,,
\eeq
which is nothing but the tidal mass-type quadrupole generalizing that of
Newtonian gravity (up to a factor $1/c^2$). Next, we directly replace the spin
tensors with their corresponding spin covectors in Eq.~\eqref{eq:sp1}, hence:
\beq
\frac{\uD \tilde{S}_{\alpha}}{\ud \tau} = 
\frac{1}{2}H^{\g\b}\ua \frac{\tilde{S}_{\g}\tilde{S}_{\b}}{m\, c} - 
\k\, \lcs \uB G^{\m\s}\frac{\tilde{S}_{\s}\tilde{S}^{\n}}{m\, c} + \OSSS \,.
\label{eq:sp2}
\eeq
Finally, after setting
\beq\label{eq:defOmegatilde}
\tilde{\Omega}_{\a\b} = \frac{\tilde{S}_{\l}}{m\, c}\left[u_{[\a}
  H_{\b]}^{\,\,\,\,\l} - \k\, \epsilon_{\a\b\m\n} u^{\m} G^{\n\l}\right] \,,
\eeq
the spin precession equation for the covariant spin
vector takes the form
\beq
\frac{\uD \tilde{S}_{\alpha}}{\ud \tau} = 
\tilde{\Omega}_{\a\b}\tilde{S}^{\b} + \OSSS \,.
\eeq
The anti-symmetric tensor $\tilde{\Omega}_{\a\b}$ may be interpreted as a
spin-precession frequency tensor.

It remains to construct a spin 3-vector $S^i$ with conserved Euclidean norm. A
``canonical'' construction is already explained in Section~2.1 of
Ref.~\cite{BMFB13}, to which the reader may refer for further details. The
precession vector governing the evolution of $S^i$ differs from that of
Ref.~\cite{BMFB13}, derived in the SO approximation, by additional terms that
are quadratic in spins.

The passage to spin 3-vectors is achieved by introducing a direct orthonormal tetrad
$e^{\,\,\m}_{\undl{\a}}$. The underlined index represents the vector label,
which we may be viewed as the tetrad index, spacetime indices being represented
by Greek letters and spatial indices by Latin letters as usual. Posing
$e^{\,\,\m}_{\undl{0}} = u^{\m}$, we see that
\beq
\tilde{S}_{\undl{0}} = \tilde{S}_{\m}e^{\,\,\m}_{\undl{0}} = \OSSS\,,
\eeq
which means that $\tilde{S}_{\undl{0}}$ may be neglected.
The squared Euclidean norm of $\tilde{S}_{\undl{a}}$ is then given by
\be 
\delta_{\undl{a}\undl{b}}\tilde{S}^{\undl{a}}\tilde{S}^{\undl{b}}=
\gamma_{\mu\nu}\tilde{S}^{\mu}\tilde{S}^{\nu} = \tilde{S}_{\mu}\tilde{S}^{\mu}
= s^{2} \,,
\ee
with $\g_{\m\n} = g_{\m\n}+u_{\m}u_{\n}$. In words, the spin vector
$\tilde{S}_{\undl{a}}$ has a conserved Euclidean norm. To define
the spin variable uniquely in some coordinate grid, we still need to specify
the choice to be made for the spatial part of the tetrad. Considering that
$\delta^{\undl{a}\undl{b}}e_{\undl{a}i}e_{\undl{b}j} = \gamma_{ij}$, a natural
choice is to take for $e_{\undl{a}i}$ the unique symmetric positive-definite square root
(in the matrix sense) of $\gamma_{ij}$.  The complete expression for the tetrad is
\beq \label{eq:tetrad}
e^{\,\,\m}_{\undl{a}}=\left(\g^{\m i} - \g^{\m 0} \frac{v^i}{c} \right) 
e_{\undl{a}i}\,.
\eeq
with  $v^{\mu} \equiv =c\, u^{\mu}/u^{0}$ denoting the coordinate velocity.
After projection on the basis vectors~\eqref{eq:tetrad},
the precession equation for the spin vector becomes
\beq
\frac{{\mathrm d}\tilde{S}_{\undl{\a}}}{{\mathrm d}\t} = 
\left(\tilde{\omega}_{\undl{\a}\undl{\b}} + 
  \tilde{\Omega}_{\undl{\a}\undl{\b}} \right) \tilde{S}^{\undl{\b}}\,,
\eeq
where we have introduced the rotation coefficients for the tetrad
\beq
\tilde{\omega}_{\undl{\a}\undl{\b}}  = -e_{\undl{\a}}^{\,\,\,\m}\,
\frac{{\mathrm D}e_{\undl{\b}\m}}{{\mathrm d}\t} \,,
\eeq
and where $\tilde{\Omega}_{\undl{\a}\undl{\b}} =
\tilde{\Omega}_{\m\n}e_{\undl{\a}}^{\,\,\,\m}\,e_{\undl{\b}}^{\,\,\n}$. Now,
as ${\mathrm d}/{{\mathrm d}\t} = u^0 {\mathrm d}/{{\mathrm d}t}$, it is
convenient to define an anti-symmetric precession frequency tensor associated
with the coordinate time as
\be \label{eq:Omegafordt}
\Omega_{\undl{\a}\undl{\b}} = \frac{1}{u^{0}} 
\left( \tilde{\omega}_{\undl{\a}\undl{\b}} + 
  \tilde{\Omega}_{\undl{\a}\undl{\b}} \right) \,.
\ee
Since $\tilde{S}^{\undl{0}}$ is negligible, the precession equation
reduces to 
\be
\frac{{\mathrm d}\tilde{S}_{\undl{i}}}{{\mathrm d}t} = 
\Omega_{\undl{i}\undl{j}} \tilde{S}^{\undl{j}} + \OSSS \,.
\ee
Moreover, from the equality $e_{\undl{0}}^{\ph{0}\mu} = u^{\mu}$, it follows
that the first term on the right-hand side of Eq.~\eqref{eq:defOmegatilde}
vanishes when projected on spatial tetrad indices, so that
\be
\tilde{\Omega}_{\undl{i}\undl{j}} = - \kappa\,
\varepsilon_{\undl{i}\undl{j}\undl{k}}
G^{\undl{k}\undl{l}}\frac{\tilde{S}_{\undl{l}}}{m\, c} \,,
\ee	
where $\varepsilon_{\undl{i}\undl{j}\undl{k}}$ or $\varepsilon_{ijk}$
(indifferently) denote
the Euclidean Levi-Civita symbol, with normalization
$\varepsilon^{123} = \varepsilon_{123} = 1$, which is linked to the four
dimensional Levi-Civita tensor by the relation
$\epsilon_{\undl{0}\undl{i}\undl{j}\undl{k}} = \varepsilon_{\undl{i}\undl{j}\undl{k}}$.

In the rest of the paper, we shall use a conserved Euclidean spin vector
$\boldsymbol{S}$ with spatial components $S^i$ in harmonic coordinates such
that
\be
S^{i} \equiv \tilde{S}^{\undl{i}} \,.
\ee
Because of the anti-symmetric character of $ \Omega_{\undl{i}\undl{j}} $, we
can finally rewrite the precession equation in terms of a precession vector $\Omega^{i} =
-\varepsilon^{\undl{i}\undl{j}\undl{k}}\Omega_{\undl{j}\undl{k}}/2$ as
\beq
\frac{{\mathrm d}S_{i}}{{\mathrm d}t} = \varepsilon_{ijk} \Omega^{j}
S^{k}+ \OSSS \,.
\eeq
It is the above precession vector $\Omega^{i}$ that will be computed, along with the
equation of motion, in Section~\ref{sec:PNdynamics}. Our results will be
displayed either in terms of the vector $S^{i}$ or of the spatial components of the
spin tensor $S^{ij}$.

\subsection{Application to self-gravitating binary systems}
\label{subsec:selfgrav}

Although the evolution equations~\eqref{eq:evolution} originally obtained by
Dixon are only suitable to describe the dynamics of test particles, their
rederivation based on the method of Tulczyjew or the Lagrangian approach of
Bailey \& Israel, regarded as effective field schemes, holds for
self-gravitating $N$ point-like body systems. Nonetheless, the validity of the
point particle model breaks down at UV scales where the post-Newtonian
expansion cannot be applied, i.e.\ for $r_A \sim R_A$, with $r$ being the
distance between the particle representing the body $A$ and the field point
$\boldsymbol{x}$. In particular, some infinities arise when computing the
gravitational field iteratively due to divergences at the particle positions
$\boldsymbol{y}_A$. The situation is even worse as we make $\boldsymbol{x}$
tend towards $\boldsymbol{y}_A$.

As usual, those infinities are cured thanks to dimensional regularization, which
preserves the invariance under diffeomorphism of general relativity, combined
with some renormalization procedure. For an appropriate choice of the space
dimension $d$, the field remains weak near $r_A = 0$ and can be computed
perturbatively in the post-Newtonian approximation. We are confident that this
leads to the correct PN dynamics because: (i) the result for the acceleration
is unambiguous up to the order 3.5PN for binaries of spinning compact objects,
(ii) it is equivalent to that obtained from other methods (see the review
paper~\cite{Bliving} for references), and notably from the approach \`a la
Einstein-Infeld-Hoffmann used by Itoh~\cite{Itoh04} in the case of spinless
bodies where no regularization is needed. Those cautions being taken, a
self-gravitating system of $N$ spinning bodies endowed with a quadrupolar
structure may be modeled by means of the following effective stress-energy
tensor, which generalizes that of Eq.~\eqref{eq:Tmunuquadrupole}:
\begin{align} \label{eq:Tmunuquadrupolebinary}
  T^{\mu\nu} &= \sum_{A=1,2} \left[ n_A \left( p_A^{(\mu}u_A^{\nu)} c +\frac{1}{3}
    R^{(\mu}_{\ph{\mu}\lambda\rho\sigma}J_A^{\nu)\lambda\rho\sigma}c^2 \right)
  \right. \nn \\
  & \qquad \quad \left. - \nabla_{\rho} 
    \left(n_A  S_A^{\rho(\mu}u_A^{\nu)} \right) -
  \frac{2}{3} \nabla_{\rho}\nabla_{\sigma} \left( n_A c^2
    J_A^{\rho(\mu\nu)\sigma} \right) \right]\,,
\end{align}
where the subscripts $A$ indicate the particle label.

The presence of poles $\propto \varepsilon^{-k} $ in the metric at a given
post-Newtonian order, with $\varepsilon \equiv d-3$ and $k$ being a positive
integer, may generate contributions in the source for the next order that
could not be recovered by resorting to a purely three dimensional
regularization. However, in the absence of such subtleties, the so-called pure
Hadamard-Schwartz regularization~\cite{BI04mult} is sufficient to get the
correct result. This prescription essentially relies on a specific use of the
Hadamard partie finie regularization, which we shall briefly discuss now (the
reader will find more details in Ref.~\cite{BF00reg}).

Let us consider a function $F(\boldsymbol{x})$ with the same regularity
properties as those arising in our problem, i.e\ smooth everywhere except at
some singular points $\boldsymbol{y}_A$ ($A=1,2,...,N$) in the neighborhood of
which its admits an expansion of the form
\be
F(\boldsymbol{x}) = \sum_{p_0 \le p \le P} r_A^p
\underset{A}{f_p}(\boldsymbol{n}_A) + o(r_A^P) 
\ee
for any integer $P$, with $\boldsymbol{n}_A =
(\boldsymbol{x}-\boldsymbol{y}_A)/r_A$. Such a function is said to be of class
$\mathcal{F}$. Its Hadamard partie finie $(F)_A$ is then defined as the
angular average of the finite part ${}_A f_0(\boldsymbol{n}_A)$:
\be
(F)_A = \int \frac{\ud \Omega_A}{4\pi}
\underset{A}{f_0}(\boldsymbol{n}_A) \,,
\ee
where $\ud \Omega_A$ denotes the elementary solid angle with direction
$\boldsymbol{n}_A$ centered on $\boldsymbol{y}_A$. The operation of taking the
Hadamard partie finie is not distributive with respect to multiplication in
the sense that, for another generic function $G(\boldsymbol{x})$ of class
$\mathcal{F}$, $(F)_A (G)_A \neq (FG)_A$ in general. Moreover, it does not
respect the Lorentz invariance. Because of the first of those two unpleasant
features, the so-defined regularization is fundamentally ambiguous as such.
Howbeit, it can still be used in practical computations provided it is
supplemented by some additional prescription. In the PNISH approach, the
post-Newtonian metric is constructed iteratively with the help of PN
potentials. Those are elementary bricks satisfying a wave-type equation (more
details are provided in Section~\ref{sec:PNdynamics}). A convenient
prescription is to define the value of a product $F G$ of two potentials (or
potential derivatives) evaluated at point $\boldsymbol{y}_A$ as $(F)_A (G)_A$.
Similarly, the regularized product of a potential $F$ and an arbitrary smooth
function $\alpha(\boldsymbol{x})$ will be given by $\alpha(\boldsymbol{y}_A)
F_A$.

Divergent integrals are cured by applying another kind of Hadamard partie
finie regularization. The regularized value of an integral with
class-$\mathcal{F}$ integrand is calculated in three main steps: (i) balls of
radius $\eta$ centered on the singular points are extracted from the
integration domain; (ii) terms that diverge near $\eta= 0$ are removed; (iii)
one goes to the limit $\eta \to 0$. The singularities that generate poles in
dimensional regularization produce logarithmic divergences in the Hadamard
one. Those are associated with cutoff parameters $s_A$ entering terms such as
$\ln (\eta/s_A)$. For consistency between the two kinds of Hadamard
regularizations, all derivatives must be evaluated in the sense of
distributions~\cite{BF00reg}. The action of the three dimensional Dirac delta
$\delta_A \equiv \delta^3(\boldsymbol{x}-\boldsymbol{y}_A)$ on test functions
must also be generalized to $\mathcal{F}$-class functions by posing $F\,
\delta_A = (F)_A \, \delta_A$.

In this context, the pure Hadamard-Schwartz regularization is an ensemble of
prescriptions designed to yield results that are ``as close as
possible'' to those obtained through dimensional regularization.
Those prescriptions demand: (i) to evaluate monomials of the form
$\alpha(\boldsymbol{x}) F_1 ... F_n$, where $\alpha(\boldsymbol{x})$ is a
smooth function and the $F_k$'s are (derivatives of) $\mathcal{F}$-class
potentials, as $\alpha(\boldsymbol{y}_A)(F_1)_A ... (F_n)_A$; (ii) to evaluate
divergent integrals by means of the Hadamard partie regularization for
integrals; (iii) to extend the definition of $\delta_A$ as explained above;
(iv) to compute all derivatives in the sense of Schwartzian distributions.

The absence of logarithmic cut-offs in the SS piece of the metric up to the
order 3PN suggests that dimensional regularization may safely be swapped for
the pure Hadamard-Schwartz one at this accuracy level. The insensitivity of
the calculations to the choice of regularization procedure has been checked
explicitly by evaluating source terms of the type $F G\, \delta_A$ in the
stress-energy tensor as $(F)_A(G)_A\, \delta_A$, thus violating the pure
Hadamard-Schwartz prescription. The results have always turned out to be
unaffected by such modifications.


\section{Next-to-leading order contributions to the post-Newto\-nian
evolution} \label{sec:PNdynamics}

We now turn to the computation of the dynamics of a binary system in the
post-Newtonian approximation, at next-to-leading order for the
quadratic-in-spin effects, i.e.\ at order $1/c^6$ (or 3PN) in the equations of
motion and at order $1/c^5$ in the equations of precession. We will recover
the results for the dynamics obtained in the
ADM~\cite{SHS07a,SHS07b,SHS08,HSS10,Steinhoff11} and
EFT~\cite{Porto06,PR06,PR08a,PR08b,Levi08,Levi12} approaches, and extend them
towards the completion of the calculation of the gravitational waves energy
flux.

We start with some general definitions in Section~\ref{subsec:PNdefinitions}.
Next, we introduce a set of potentials parametrizing the PN metric in
Section~\ref{subsec:sigmaeompot}, and express the quantities of interest in
terms of these potentials. In Section~\ref{subsec:PNpotentials}, we present
their computation, and finally in Section~\ref{subsec:resultseom} the results
obtained for the dynamics as well as various tests of their correctness. The
lengthier calculations are all performed by means of the algebraic computing
software Mathematica\textregistered{} supplemented by the tensor calculus
package xAct~\cite{xtensor}.


\subsection{General definitions} \label{subsec:PNdefinitions}

The two objects are represented as quadrupolar point particles as explained
above. An important ingredient of the formalism is the treatment of the
infinite self-field of the point particles, essentially represented by means
of Dirac deltas, through the pure Hadamard-Schwartz regularization procedure
discussed in Section~\ref{subsec:selfgrav}. The distributional contributions
yielded by derivatives are handled by using the Gel'fand-Shilov
formula~\cite{Gelfand}. We found that at this order in spin, we have to keep
track of distributional contributions in the metric itself to obtain the
correct result for the wave generation formalism, as will be detailed in
Section~\ref{subsec:moments}.

The general structure of the equations of motion and precession is as follows:
\begin{subequations}
\begin{align}\label{eq:PNstructA}
  \bm{A} &= \bm{A}_{\rm NS}^{\rm N} + 
  \frac{1}{c^{2}} \bm{A}_{\rm NS}^{\rm 1PN} + 
  \frac{1}{c^{4}} \bm{A}_{\rm NS}^{\rm 2PN} + 
  \frac{1}{c^{5}} \bm{A}_{\rm NS}^{\rm 2.5PN} + 
  \frac{1}{c^{6}} \bm{A}_{\rm NS}^{\rm 3PN} \nn\\
  & \quad + \frac{1}{c^{3}} \bm{A}_{\rm SO}^{\rm N} + 
  \frac{1}{c^{5}} \bm{A}_{\rm SO}^{\rm 1PN} + 
  \frac{1}{c^{4}} \bm{A}_{\rm SS}^{\rm N} + 
  \frac{1}{c^{6}} \bm{A}_{\rm SS}^{\rm 1PN} + \calO(7) \,, \\ \label{eq:PNstructO}
  \bm{\Omega} &= \frac{1}{c^{2}}\bm{\Omega}_{\rm NS}^{\rm N} + 
  \frac{1}{c^{4}}\bm{\Omega}_{\rm NS}^{\rm 1PN} + 
  \frac{1}{c^{3}}\bm{\Omega}_{\rm SO}^{\rm N} + 
  \frac{1}{c^{5}}\bm{\Omega}_{\rm SO}^{\rm 1PN} + \calO(6) \,,
\end{align}
\end{subequations}
where the spin order in Eq.~\eqref{eq:PNstructO} indicates the contribution in
$\bm{\Omega}$ itself, rather than in $\dot{\bm{S}} =\bm{\Omega}\times\bm{S}$
(notably the SO terms feature the constants $\kappa_{1,2}$ and actually correspond
to SS terms in $\dot{\bm{S}}$). The 2.5PN NS terms in the
acceleration are the first manifestation of radiation reaction.

We use the same notations as in previous works. Three-dimensional indices are
represented with Latin letters $a,b,...$ or $i,j,...$, and are risen or
lowered with the Euclidean metric $\delta_{ij}$; we do not distinguish between
upper and lower indices. We sometimes use boldface for Euclidean vectors. The
positions and velocities of the two bodies are denoted by $y_{1}^{i}$,
$y_{2}^{i}$ and $v_{1}^{i}$, $v_{2}^{i}$. Apart from the separation distance
$r_{12} = |\bm{y}_{12}| = |\bm{y}_{1} - \bm{y}_{2}|$ which we have already defined,
we shall need the separation direction $n_{12}^{i} = (y_{1}^{i} -
y_{2}^{i})/r_{12}$. The symbol $1 \leftrightarrow 2 $ indicates the same
expression as the one before it, with the label of the two particles exchanged.
The results are expressed in terms of the spatial components $S_{1}^{ij}$,
$S_{2}^{ij}$ of the spin tensor $S^{\mu\nu}$, as well as the spin vectors
$S_{1}^{i}$, $S_{2}^{i}$ of conserved Euclidean norm as defined above, in
Section~\ref{subsec:defspinvector}. The mixed components $S^{0i}$ of the spin
tensors can always be eliminated with the help of the spin supplementary
condition~\eqref{eq:ssc}. We allow repeated indices in scalars quantities
enclosed by parenthesis, in the absence of a risk of confusion.

In harmonic (or DeDonder) gauge, the gravitational field equations can be
rewritten as
\be \label{eq:Einsteinharmonic}
\Box h^{\mu\nu} = \frac{16\pi G}{c^{4}} |g| T^{\mu\nu} + 
\Lambda^{\mu\nu}\left[ h \right] \equiv \frac{16\pi G}{c^{4}} \tau^{\mu\nu} \,,
\ee
where the stress-energy pseudo-tensor $\tau^{\mu\nu}$ includes both matter and
field contributions through $T^{\mu\nu}$ and $\Lambda^{\mu\nu}$, the latter
source term being at least quadratic in $h^{\mu\nu}$. The field
equations~\eqref{eq:Einsteinharmonic}, when iterated order by order, yield
a solution expressed formally in terms of a hierarchy of potentials of increasing
complexity and post-Newtonian order (see Refs.~\cite{PB02,BFN05} for the
precise definition of this iteration in the near-zone).

Since on the one hand we are working at order $1/c^6$ in the equations of
motion, and on the other hand the spin contributions always come at relative
$1/c$ order at least, only the so-called 2PN metric and
potentials (i.e.\ necessary for the 2PN non-spinning case) are required. In fact, we
will see below that, among the potentials arising at the order 2PN, only
$\hat{X}$ turns out to be needed. For completeness, we
quote here the result of the iteration for the 2PN metric,
which reads
\begin{subequations}\label{eq:metricg}
\begin{align} 
  g_{00} &=  -1 + \frac{2}{c^{2}}V - \frac{2}{c^{4}} V^{2} + \frac{8}{c^{6}}
  \left(\hat{X} + V_{i} V_{i} + \frac{V^{3}}{6}\right) +\calO(8)\,,\\ 
  g_{0i} & = - \frac{4}{c^{3}}
  V_{i} - \frac{8}{c^{5}} \hat{R}_{i} + \calO(7)\,,\\ 
  g_{ij} & = \delta_{ij} \left[1 +
    \frac{2}{c^{2}}V + \frac{2}{c^{4}} V^{2} \right] + 
  \frac{4}{c^{4}}\hat{W}_{ij} + \calO(6) \,.
\end{align}
\end{subequations}\noindent
The potentials therein are defined as\footnote{Possible contributions to the metric of
  non-linear tail terms, which are not made of (products of) elementary
  potentials defined by means of the operator $\Box_{\mathcal{R}}^{-1}$, do not
  arise below the order 4PN~\cite{PB02,BFN05}.}
\begin{subequations}\label{eq:defpotentials}
\begin{align}
  V & = \Box_{\mathcal{R}}^{-1}[-4 \pi G\, \sigma]\;,\label{V} \\ 
  V_{i} &= \Box_{\mathcal{R}}^{-1}[-4 \pi G\, \sigma_{i}]\,, \\
  \hat{X} &= \Box_{\mathcal{R}}^{-1}\left[\vphantom{\frac{1}{2}} -
    4 \pi G\, V \sigma_{ii} + \hat{W}_{ij}\partial_{ij} V + 
    2 V_{i} \partial_{t} \partial_{i} V + V \partial_{t}^{2}V\right. \nonumber \\
  & \qquad \qquad \left. + \frac{3}{2}(\partial_{t} V)^{2} - 
    2\partial_{i} V_{j}\partial_{j} V_{i}\right] \,, \\
  \hat{R}_{i} & = \Box_{\mathcal{R}}^{-1}\left[-4 \pi G\, 
    (V \sigma_{i} - V_{i} \sigma) - 2\partial_{k} V \partial_{i} V_{k} - 
    \frac{3}{2} \partial_{t} V \partial_{i} V\right]\,, \\ 
  \hat{W}_{ij} & =  \Box_{{\cal R}}^{-1}\left[-4 \pi G\, 
    (\sigma_{ij} - \delta_{ij} \sigma_{kk}) - 
    \partial_{i} V \partial_{j}V\right]\,, \label{Wij} 
\end{align}
\end{subequations}
where the $\sigma$, $\sigma_{i}$, $\sigma_{ij}$ quantities are convenient
matter source densities defined as
\begin{equation}\label{eq:defsigma}
  \sigma=\frac{1}{c^{2}}(T^{00}+T^{ii}) \,, \quad 
  \sigma_{i}=\frac{1}{c}T^{0i} \,, \quad \sigma_{ij}=T^{ij} \,,
\end{equation}
while $\Box_{\mathcal{R}}^{-1}$ stands for the PN-expanded retarded
d'Alembertian operator acting on a function $f(\boldsymbol{x},t)$ as~\cite{PB02,BFN05}
\begin{equation}\label{eq:defdalembertian}
  (\Box_{\mathcal{R}}^{-1}f)(\boldsymbol{x},t) = 
  - \frac{1}{4\pi} \sum_{n\geq 0} \frac{(-1)^{n}}{n!} 
  \mathrm{FP}_{B=0} \int \ud^{3}x' (|\boldsymbol{x}'|/r_{0})^{B} 
  |\boldsymbol{x}-\boldsymbol{x}'|^{n-1}
  f^{(n)}(\boldsymbol{x}',t)  \,. 
\end{equation}
Here $\mathrm{FP}_{B=0}$ denotes the so-called Finite Part regularization, and
$r_{0}$ is an associated arbitrary length scale. This regularization is used
and described in Section~\ref{subsec:mpnformalism} for the wave generation
formalism, but in Eq.~\eqref{eq:defdalembertian} it cures the divergences of
the near-zone post-Newtonian metric at infinity rather than the divergences
of the multipolar far-zone expansion at the origin. At the order we are
considering here, it does not matter for the equations of motion. In
particular the final results are all independent of the scale
$r_{0}$\footnote{At the 3PN non-spinning order, the scale $r_{0}$ does appear
  in the final results for the dynamics, but it disappears when considering
  gauge-invariant expressions such as $E(\omega)$, the conserved energy as a
  function of the orbital frequency.}.


\subsection{Matter source and equations of motion in terms of 
potentials} \label{subsec:sigmaeompot}

In this section, we introduce convenient additional definitions for the matter
source in the PN context. In the covariant
expression~\eqref{eq:Tmunuquadrupolebinary}, the worldline integration
contained in the particle densities $n_A$ [see Eq.~\eqref{eq:scalar_density}]
can be performed explicitly in a definite coordinate grid $(t,\bm{x})$. This
results in
\begin{align}
  T^{\alpha\beta} = \sum_{A=1,2} \bigg[U_A^{\alpha\beta} \delta_A + \nabla_\mu
  \Big(U_A^{\alpha\beta\mu} \delta_A \Big) + \nabla_\mu \nabla_\nu
  \Big(U_A^{\alpha\beta\mu\nu} \delta_A \Big)
\bigg] \,,
\end{align}
where we have defined
\begin{subequations}\label{eq:TmunuU}
\begin{align}
  &U_A^{\alpha\beta} = \frac{1}{u_A^0 \sqrt{-g}} \Big(c \, p_A^{(\alpha}u_A^{\beta)} +
  \frac{1}{3} R^{(\alpha}_{A\,\lambda\mu\nu} J_A^{\beta)\lambda\mu\nu}\Big)\,,\\
  &U_A^{\alpha\beta\mu} = 
  \frac{1}{u_A^0 \sqrt{-g}}  u_A^{(\alpha}S_A^{\beta)\mu}\,,\\
  &U_A^{\alpha\beta\mu\nu} =  \frac{1}{u_A^0 \sqrt{-g}} \Big(-\frac{2}{3}c^2 
  J_A^{\mu(\alpha\beta)\nu}\Big) \,.
\end{align}
\end{subequations}
Here $u^{0} = 1/\sqrt{-g^{A}_{\mu\nu}v^{\mu}v^{\nu}/c^{2}}$, $v^{\mu} =
(c,v^{i})$ (so that $u^{\mu} = u^{0}v^{\mu}/c$), and the label index $A$ on
metric-dependent quantities means that they are to be regularized
according to Hadamard regularization at the location of the particle $A$. In
terms of partial derivatives, we have
\begin{align}
  T^{\alpha\beta} = \sum_{A=1,2} \bigg[\mathcal{T}_A^{\alpha\beta} \delta_A + 
  \frac{1}{\sqrt{-g}}\partial_\mu 
  \Big(\mathcal{T}_A^{\alpha\beta\mu} \delta_A \Big) + \frac{1}{\sqrt{-g}}
  \partial_{\mu\nu} \Big(\mathcal{T}_A^{\alpha\beta\mu\nu} \delta_A \Big)
  \bigg] \,,
\end{align}
with
\begin{subequations}
\begin{align}
  &\mathcal{T}_A^{\alpha\beta} = U_A^{\alpha\beta} + 2
  \Gamma^{(\alpha}_{A\,\mu\nu} U_A^{\beta)\mu\nu} + (\partial_\lambda
  \Gamma^{(\alpha}_{\mu\nu})_{A}\, U_A^{\beta)\lambda\mu\nu}
  \nonumber \\ & \qquad + \Gamma^{(\alpha}_{A\,\rho\lambda} (U_A^{\beta)\lambda\mu\nu}
  \Gamma^{\rho}_{A\,\mu\nu} - \Gamma^{\beta)}_{A\,\mu\nu}
  U_A^{\rho\lambda\mu\nu})\,,\\
  &\mathcal{T}_A^{\alpha\beta\mu} =  \sqrt{-g} (U_A^{\alpha\beta\mu} +
  \Gamma^{\mu}_{A\,\nu\lambda} U_A^{\alpha\beta\nu\lambda} - 
  2 \Gamma^{(\alpha}_{A\,\nu\lambda} U_A^{\beta)\mu\nu\lambda})\,,\\
  &\mathcal{T}_A^{\alpha\beta\mu\nu} =  \sqrt{-g} U_A^{\alpha\beta\mu\nu} \,.
\end{align}
\end{subequations}
By using Eqs.~\eqref{eq:TmunuU}, and the definitions of $\sigma$,
$\sigma_{i}$, $\sigma_{ij}$ given in Eqs.~\eqref{eq:defsigma}, we arrive at the
following expressions in terms of metric potentials
\begin{subequations}\label{eq:sigmapot}
\begin{align}  \label{eq:sigmapots}
  \sigma &= m_1 \delta_1 \bigg[
  1 + \frac{1}{c^{2}} \Big(\frac{3}{2} v_{1}^2
  -  V\Big) + \frac{1}{c^{4}} \Big(\frac{7}{8} v_{1}^4
  - 4 \bigl(v_{1}^{a} V_{a}\bigr)
  - 2 \hat{W} + \frac{1}{2} v_{1}^2 V
  + \frac{1}{2} V^2\Big) \nonumber\\
  & \qquad + \frac{1}{m_1 c^{5}} \Big(2 (S_{1}^{ab} v_{1}^{a} \partial_{b}V)
  - 4 (S_{1}^{ab} \partial_{b}V_{a})\Big) + 
  \frac{1}{c^{6}} \Big(-8 (\hat{R}_{a} v_{1}^{a})
  + \frac{11}{16} v_{1}^6 \nonumber\\ & \qquad \qquad 
  - 10 v_{1}^{2} (v_{1}^{a} V_{a})
  - 4 (V_{a} V_{a}) + 2 (v_{1}^{a} v_{1}^{b} \hat{W}_{ab}) - 3 v_{1}^2 \hat{W}
  - 8 \hat{Z} + \frac{33}{8} v_{1}^4 V
  \nonumber\\ & \qquad \qquad - 4 (v_{1}^{a} V_{a}) V
  + 2 \hat{W} V + \frac{11}{4} v_{1}^2 V^2 -  \frac{1}{6} V^3
  - 4 \hat{X} -  \frac{\kappa_1}{2m_1^2}  
  (S_{1}^{ba} S_{1}^{bi} \partial_{ia}V)\Big) \bigg]\nonumber\\
  & \qquad + \frac{1}{\sqrt{- g}}\partial_{k}\bigg\{ \delta_1 \bigg[ - 
  \frac{2 S_{1}^{ka} v_{1}^{a}}{c^3} + \frac{S_{1}^{ka}}{c^5} 
  \Big(-4 V v_{1}^{a} + 4 V_{a}\Big) \nonumber\\ & \qquad \qquad + 
  \frac{\kappa_1}{m_1 c^6} \Big( S_{1}^{ab} S_{1}^{kb} \partial_{a}V -
  (S_{1}^{ab} S_{1}^{ab}) \partial_{k}V\Big)\bigg]\bigg\}\nonumber\\
  & \qquad  + \frac{\kappa_1}{2 m_1 c^6 \sqrt{- g}}\partial_t^{2} 
  \Big(\delta_1 S_{1}^{ab} S_{1}^{ab}\Big) + 
  \frac{\kappa_1}{m_1 c^6 \sqrt{- g}} \partial_t 
  \partial_{k} \bigg[\delta_1 \Big(- S_{1}^{ab} S_{1}^{kb} v_{1}^{a} + 
  \bigl(S_{1}^{ab} S_{1}^{ab}\bigr) v_{1}^{k}\Big) \bigg]
  \nonumber\\
  & \qquad + \frac{\kappa_1}{m_1\sqrt{- g}} \partial_{kl}
  \bigg\{\delta_1 \bigg[\frac{S_{1}^{ka} S_{1}^{la}}{2 c^4} + 
  \frac{1}{c^6}\Big(S_{1}^{ka} S_{1}^{la} \big(\frac{3}{4} 
  v_{1}^2 + \frac{3}{2} V\big) - 
  \frac{1}{2} S_{1}^{ka} S_{1}^{lb} v_{1}^{a} v_{1}^{b} \nonumber\\ & \qquad
  \qquad \qquad + 
  \frac{1}{2} (S_{1}^{ab} S_{1}^{ab}) v_{1}^{k} v_{1}^{l} + 
  S_{1}^{ab} (- S_{1}^{lb} v_{1}^{a} v_{1}^{k} -  
  S_{1}^{kb} v_{1}^{a} v_{1}^{l})\Big)\bigg] \bigg\} \nonumber\\ &
  + 1 \leftrightarrow 2 
  + \mathcal{O}\Big(\frac{1}{c^7}\Big)\,, \\
  \sigma_i &= m_1 \delta_1 \bigg[ v_{1}^{i}
  + \frac{1}{c^{2}} \Big(\frac{1}{2} v_{1}^2
  v_{1}^{i}-  V v_{1}^{i}\Big) -  \frac{ S_{1}^{ia} \partial_{a}V}{2m_1c^{3}}
  \nonumber\\ & \qquad + \frac{1}{c^{4}} \Big(\frac{3}{8} v_{1}^4 v_{1}^{i} 
  - 4 (v_{1}^{a} V_{a}) v_{1}^{i}
  - 2 \hat{W} v_{1}^{i}
  + \frac{3}{2} v_{1}^2 V v_{1}^{i} + \frac{1}{2} V^2
  v_{1}^{i}\Big) \bigg]
  \nonumber\\ & \qquad 
  +\frac{1}{2 c^3 \sqrt{- g}}\partial_t \Big(\delta_1 S_{1}^{ia} v_{1}^{a}\Big)+
  \frac{1}{\sqrt{- g}} \partial_{k}\bigg[\delta_1 \Big(\frac{S_{1}^{ik}}{2 c} - 
  \frac{S_{1}^{ka} v_{1}^{a} v_{1}^{i}}{2 c^3}\Big)\bigg] \nonumber\\
  & \qquad - \frac{\kappa_1}{2 m_1 c^4
    \sqrt{- g}}\partial_t \partial_{k}\Big(\delta_1
  S_{1}^{ia} S_{1}^{ka}\Big)
  \nonumber\\ & \qquad + \frac{\kappa_1}{m_1 c^4\sqrt{-
      g}}\partial_{kl}\bigg[\delta_1 \Big(\frac{1}{2} S_{1}^{ka} S_{1}^{la}
  v_{1}^{i} - 
  \frac{1}{4} S_{1}^{ia} S_{1}^{la} v_{1}^{k} -  \frac{1}{4} S_{1}^{ia}
  S_{1}^{ka} v_{1}^{l}\Big) \bigg]  \nonumber\\ & + 1 \leftrightarrow 2 + 
  \mathcal{O}\Big(\frac{1}{c^5}\Big)\,, \\
  \sigma_{ij}&= m_1\delta_1\bigg[
  v_{1}^{i} v_{1}^{j}
  + \frac{1}{c^{2}}  
  \Big(\frac{1}{2} v_{1}^2 v_{1}^{i} v_{1}^{j}
  -  V v_{1}^{i} v_{1}^{j}\Big)\bigg]\nonumber \\ & \qquad  +\frac{1}{c\sqrt{-
      g}}\partial_{k}\bigg[\delta_1 
  \Big(\frac{1}{2} S_{1}^{jk} v_{1}^{i}+ \frac{1}{2}
  S_{1}^{ik} v_{1}^{j}\Big)\bigg]  +  1 \leftrightarrow 2 + 
  \mathcal{O}\Big(\frac{1}{c^3}\Big) \,.
\end{align}
\end{subequations}
Here we have dropped the indices on the metric potentials, as we recall that,
according to the pure Hadamard-Schwartz regularization, we can indifferently
consider any quantity in factor of a Dirac delta as regularized, according to
the rule $F \delta_{1} = (F)_{1}\delta_{1}$. Notice however that the
$1/\sqrt{-g}$ prefactors must \textit{not} be evaluated at point
$\boldsymbol{y}_1$, i.e.\ they are still functions of the field point
$\bm{x}$. These matter sources display explicit factors with spins and other
without spin, but we should always keep in mind that there are secondary spin
contributions coming from the potentials themselves.

Let us now turn to the expression of the equations of evolution in terms of
the metric potentials~\eqref{eq:defpotentials}. If we pose $P_{\mu} =
p_{\mu}/m$, using $\ud /\ud \tau = u^{0} \ud / \ud t$, the covariant
equations of motion~\eqref{eq:dpdt} may be put in the form
\be\label{eq:dPdtF}
\frac{\ud P_{\mu}}{\ud t} = F_{\mu} \equiv
\Gamma^{\rho}_{\ph{\rho}\mu\nu}v^{\nu}\frac{p_{\rho}}{m} - 
\frac{1}{2 m c}R_{\mu\nu\rho\sigma} v^{\nu}S^{\rho\sigma} - 
\frac{c^{2}}{6  m u^{0}} 
J^{\lambda\nu\rho\sigma}\nabla_{\mu}R_{\lambda\nu\rho\sigma} + 
\calO(S^{3}) \,,
\ee
where $P_{\mu} = p_{\mu}/m$ can be read from Eq.~\eqref{eq:pmutoumu}. For a lower
spatial index $\mu=i$, we decompose $P_{\mu}$ as (coming back to notations for the body
1)
\begin{align}
  P_{1i} &= P_{1i}^{\rm NS} + P_{1i}^{\rm SS} \,, \nn\\
  F_{1i} &= F_{1i}^{\rm NS} + F_{1i}^{\rm SO}+ F_{1i}^{\rm SS} \,.
\end{align}
Here, the order in spin refers to the order in the spin tensor as it reads in
the formulas~\eqref{eq:pmutoumu} and~\eqref{eq:dpdt}, but we should recall
that there are also spin contributions coming from the potentials themselves,
as well as from the replacement of accelerations using the equations of
motion. We see from Eq.~\eqref{eq:pmutoumu} that $P_{i}$ has no SO part in this
sense, and its NS part comes from $P_{i} = u^{0}g_{i\nu}v^{\nu}/c$. The NS
part of $F_{i}$ comes from the usual connexion term in the geodesic equation,
the first term in Eqs.~\eqref{eq:dPdtF}. As the NS and SO parts can already be
found e.g.\ in Eqs.~(2.12) of Ref.~\cite{NB05} and in Eqs.~(3.7) of
Ref.~\cite{MBFB13}, we only display here the SS pieces:
\begin{subequations}  \label{eq:PFpot}
\begin{align}
  P_{1i}^{(\text{S}^2)} &= \frac{1}{m_1^2 c^6} 
  \bigg\{ - S_{1}^{ab} S_{1}^{ib} \partial_t \partial_{a}V + 
  2 S_{1}^{bj} S_{1}^{ia} \partial_{ja}V_{b} + 
  S_{1}^{ib} \bigl(S_{1}^{bj} v_{1}^{a} \partial_{ja}V  - 
  2 S_{1}^{aj} v_{1}^{a} \partial_{jb}V \bigr) \nonumber\\
  & \qquad + 
  \kappa_1 \bigg[ \frac{3}{2} (S_{1}^{ai} S_{1}^{ab} \partial_{ib}V) v_{1}^{i} -  
  S_{1ab} S_{1}^{ib} \partial_t \partial_{a}V\nonumber\\
  & \qquad \qquad + (S_{1}^{ab} S_{1}^{ab}) \partial_t \partial_{i}V + 
  (S_{1}^{ab} S_{1}^{ab}) v_{1}^{a} \partial_{ia}V + 
  S_{1}^{bj} S_{1}^{ib} v_{1}^{a} \partial_{ja}V \nonumber\\
  & \qquad \qquad  + 2 S_{1}^{ab} S_{1}^{bj} v_{1}^{a} \partial_{ji}V + 
  S_{1}^{aj} S_{1}^{ab} (-2 \partial_{jb}V_{i} + 
  2 \partial_{ji}V_{b}) \bigg] \bigg\}\,, \\
  F_{1i}^{(\text{S}^2)} &= \frac{\kappa_1 }{2m_1^2 c^4} 
  S_{1}^{aj} S_{1}^{ab} \partial_{ji}\partial_{b}V + 
  \frac{\kappa_1}{m_1^2 c^6}  \bigg[\bigl(S_{1}^{ab} S_{1}^{ab}\bigr) 
  v_{1}^{a} \partial_t \partial_{ia}V \nonumber\\ & \qquad + 
  \frac{1}{2} \bigl(S_{1}^{ab} S_{1}^{ab}\bigr) 
  \partial_t^{2} \partial_{i}V -  
  \frac{3}{2} \bigl(S_{1}^{ai} S_{1}^{ab} \partial_{ib}V\bigr) \partial_{i}V + 
  \bigl(S_{1}^{ab} S_{1}^{ab}\bigr) \partial_{a}V \partial_{ia}V
  \nonumber\\ & \qquad  + 
  \frac{1}{2} \bigl(S_{1}^{ab} S_{1}^{ab}\bigr) v_{1}^{a} v_{1}^{b} 
  \partial_{iba}V -  
  S_{1}^{bj} S_{1}^{ib} \partial_{a}V \partial_{ja}V + 
  S_{1}^{ab} S_{1}^{bj} \Bigl(v_{1}^{a} \partial_t \partial_{ji}V\nonumber\\
  & \qquad \qquad + 4 \partial_{a}V \partial_{ji}V\Bigr) + 
  S_{1}^{aj} S_{1}^{ab} \Bigl(2 \partial_t \partial_{ji}V_{b} + 
  \frac{3}{4} v_{1}^{2} \partial_{jib}V + 
  \frac{1}{2} V \partial_{jib}V\Bigr) \nonumber\\
  & \qquad + 
  S_{1}^{aj} \Bigl(2 S_{1}^{jk} v_{1}^{a} v_{1}^{b} \partial_{kib} V -  
  \frac{1}{2} S_{1}^{bk} v_{1}^{a}v_{1}^{b} 
  \partial_{kji}V\Bigr)\nonumber\\ & \qquad + 
  S_{1}^{bk} S_{1}^{bj} \bigl(2 v_{1}^{a} \partial_{kia} V_{j} - 
  2 v_{1}^{a} \partial_{kji}V_{a}\bigr)\bigg] \,.
\end{align}
\end{subequations}
In these formulas, the potentials and their derivatives are to be understood as
regularized at the location of body 1.

For the equation of precession of the conserved-norm spin, we decompose
similarly the precession vector into a NS and a SO part, before replacement of
the potentials. We obtain
\begin{align}
  \Omega_{1}^i = \left( \Omega_{1}^i \right)_{\rm NS} + \left( \Omega_{1}^{i}
  \right)_{\rm SO} \,,
\end{align}
where $\left( \Omega_{1}^i \right)_{\rm NS}$ is given by
Eq.~(2.19) in~\cite{BMFB13} and where
\begin{align}\label{eq:Omegapot}
  \left( \Omega_{1}^{i} \right)_{\rm SO} &= 
  \frac{\kappa_1}{m_1 c^{3}} S_{1}^{a} \partial_{ia}V + 
  \frac{1}{m_1 c^{5}} \bigg\{v_{1}^{i} 
  \left( S_{1}^{a} v_{1}^{b} \partial_{ba}V \right) - 
  \frac{1}{2} \bigl(S_{1}^{a} v_{1}^{a}\bigr) \partial_t \partial_{i}V + 
  S_{1}^{a}v_{1}^{b}\partial_{ab}V_{i} \nonumber \\
  & - S_{1}^{a} (v_{1}^{a}v_{1}^{a}) \partial_{ia}V +
  \frac{1}{2}\left(S_{1}^{a}v_{1}^{a}\right)v_{1}^{a}\partial_{ia}V - 
  \left( S_{1}^{a}v_{1}^{a} \right) \partial_{ia}V_{a} + 
  S_{1}^{a}v_{1}^{b}\partial_{ia}V_{b} \nonumber \\
  & + \kappa_1 \bigg[- 
  \bigl(S_{1}^{a} \partial_t \partial_{a}V\bigr) v_{1}^{i} -
  \frac{3}{2} \bigl(S_{1}^{a} v_{1}^{b} \partial_{ba}V\bigr) v_{1}^{i} - 
  \bigl(S_{1}^{a} v_{1}^{a}\bigr) \partial_t \partial_{i}V \nonumber \\
  & \qquad  + S_{1}^{i} \Bigl(2 (v_{1}^{a} \partial_t \partial_{a}V) + 
  (\partial_{a}V \partial_{a}V)+ (v_{1}^{a} v_{1}^{b} \partial_{ba}V) + 
  \partial_t^{2} V\Bigr) \nonumber \\
  & \qquad  - 3 \bigl(S_{1}^{a} \partial_{a}V\bigr) \partial_{i}V - 
  \frac{3}{2} \bigl(S_{1}^{a} v_{1}^{a}\bigr) v_{1}^{a} \partial_{ia}V + 
  S_{1}^{a} \Bigl(2 \partial_t \partial_{a}V_{i} + 
  2 \partial_t \partial_{i}V_{a} \nonumber \\ \qquad \qquad
  & \qquad + 2 v_{1}^{b} \partial_{ba}V_{i} + 
  \frac{3}{2} v_{1}^2 \partial_{ia}V - 3 V \partial_{ia}V -
  4 v_{1}^{b} \partial_{ia}V_{b} + 
  2 v_{1}^{b} \partial_{ib}V_{a}\Bigr)\bigg]\bigg\} \,.
\end{align}
The contributions featuring $\kappa_{1}$ come directly from the second term
in Eq.~\eqref{eq:Omegafordt}, while the other contributions come from the first term
there. The time derivatives of the velocities that
enter the definition of the tetrad are replaced by the expression of the
acceleration in terms of potentials, which include SO terms (as given for
instance in Section~3 of~\cite{MBFB13}).


\subsection{Spin contributions in the metric
  potentials} \label{subsec:PNpotentials}

We now investigate the spin contributions to the metric potentials introduced
in Section~\ref{subsec:PNdefinitions}. As we are effectively working at the
next-to-leading order, calculating these contributions from the results already
presented above will be rather straightforward and we will only need
to resort to well-known techniques.

By inspection of the matter sources~\eqref{eq:sigmapot}, one can see that the
SO and SS contributions to the metric potentials start at the following PN
orders:\footnote{It is implicitly understood that the orders $n$ showed in the
 terms $\calO(n)$ below take their highest possible values.}
\begin{align}\label{eq:potspinorder}
  V^{\rm SO} &= \calO(3) \,, \qquad V^{\rm SS} = \calO(4) \,, \nn\\
  V_{i}^{\rm SO} &= \calO(1) \,, \qquad V_{i}^{\rm SS} = \calO(4) \,, \nn\\
  \hat{X}^{\rm SO} &= \calO(1) \,, \qquad 
  \hat{X}^{\rm SS}  = \calO(2) \,, \nn\\
  \hat{R}_{i}^{\rm SO} &= \calO(1) \,, \qquad 
  \hat{R}_{i}^{\rm SS} = \calO(4) \,, \nn\\
  \hat{W}_{ij}^{\rm SO} &= \calO(3) \,, \qquad 
  \hat{W}_{ij}^{\rm SS} = \calO(4) \,.
\end{align}
From Eqs.~\eqref{eq:PFpot} and~\eqref{eq:Omegapot},
we see that it is sufficient to compute the new SS contributions of the
potentials $V$ at the order 3PN (the leading order contribution at the order 2PN
being already known, see e.g.\ Ref.~\cite{BFH12}), $V_{i}$ at the order 2PN, and
$\hat{X}$ at the order 1PN.

We turn now to the caculation of $V^{\rm SS}$ at the 3PN order which actually
corresponds to the relative 1PN order. Truncating
Eq.~\eqref{eq:defdalembertian} appropriately, we have (dropping the
$\mathrm{FP}_{B=0}$ regularization, which plays no role for compact sources)
\be \label{eq:Vexpand}
V^{\rm SS} = \left[ \int \frac{\ud^{3}x'}{|x-x'|} \sigma(t,x') - 
  \frac{1}{c}\frac{\ud}{\ud t} \int \ud^{3}x' \sigma(t,x') + 
  \frac{1}{2c^{2}}\frac{\ud^{2}}{\ud t^{2}} 
  \int \ud^{3}x' |x-x'| \sigma(t,x') \right]_{\rm SS} +\calO(8) \,,
\ee
Because we are working at the next-to-leading order, various indirect
contributions appear. Aside from the SS terms generated directly by the SS
terms of $\sigma$ given in Eq.~\eqref{eq:sigmapots}, there are contributions from
the SO 0.5PN part of $V_{i}$ in the SO 2.5PN part of $\sigma$, from
the SS 2PN part of $V$ in the NS 2PN part of $\sigma$, and from
the acceleration replacement featuring the SS 2PN part of $a^{i}$ in
the second time derivative of the integral of the NS Newtonian part of
$\sigma$ (in the third term above).

In the following, to present the result in a more compact form, we adopt short
cut notations: for any vectors $a$, $b$ and spin tensors $S_{A,B}$,
we define the scalars $(ab) = a^{i}b^{i}$, $(S_{A}S_{B}) \equiv
S_{A}^{ij}S_{B}^{ij} $, $(a S_{A} b) \equiv a^{i}S_{A}^{ij}b^{j}$, and
$(aS_{A}S_{B}b) \equiv a^{i} S_{A}^{ik}S_{B}^{jk}b^{j}$ (beware of the convention
for the order of the indices on the spin tensors).

We replace in Eq.~\eqref{eq:Vexpand} the full expression~\eqref{eq:sigmapots}
for $\sigma$, perform integration by parts when derivatives of Dirac deltas
occur, and compute the resulting integrals using Hadamard regularization, i.e.\
$\int \ud^{3}xF \delta_{1} = (F)_{1}$. The metric potentials can be considered
as regularized when appearing in factor of a Dirac delta in the integrals,
according to the pure Hadamard-Schwartz rule~\cite{BDE04} $F \delta_{1} =
(F_{1})\delta_{1}$.

An important point is that the derivatives have to be treated in a
distributional sense. For the first time in our formalism, we have to take
into account an essential distributional term in the potential $V$ itself. The
leading order result is indeed
\be\label{eq:Vleading}
V^{\rm SS} = \frac{G \kappa_{1}}{2 c^{4} m_{1}}
S_{1}^{ik}S_{1}^{jk} \partial_{ij}\left( \frac{1}{r_{1}} \right) + \exch +
\calO(6) \,,
\ee
which, along with a non-distributional contribution, yields a distributional
term given by
\be\label{eq:Vdistr}
V^{\rm SS}_{\rm distr} = \frac{G \kappa_{1}}{2 c^{4} m_{1} r_{1}^{3}}
S_{1}^{ij}S_{1}^{ij} \frac{4\pi}{3} \delta_{1} + \exch + \calO(6) \,.
\ee
This distributional term will play no role in the derivation of the equations
of motion themselves, but it will produce a net contribution when computing the
mass source quadrupole moment, as explained below in
Section~\ref{subsec:moments}. Because, in this computation of the quadrupole moment, the
$V$ potential is only needed at the 2PN order, we will not need to
consider possible distributional terms at the higher 3PN order.

Gathering the different non-distributional contributions we obtain
\begin{align}\label{eq:Vnondistr}
  V^{\rm SS}_{\rm non \; distr} &= 
  \frac{G \kappa_{1}}{2 c^{4} m_{1} r_{1}^{3}} \left( 3 (n_{1}S_{1}S_{1}n_{1}) - 
    (S_{1}S_{1}) \right) \nn\\ 
  & + \frac{G \kappa_{1}}{c^{6} m_{1}} \left[ (S_{1}S_{1}) 
    \left( \frac{3 (n_{1}v_{1})^{2}}{4 r_{1}^{3}} - 
      \frac{(v_{1}v_{1})}{r_{1}^{3}} - 
      \frac{3 G m_{2} (n_{12}n_{1})}{4 r_{12}^{4}} + 
      \frac{3 G m_{2} (n_{12}n_{2})}{4 r_{12}^{4}} + 
      \frac{3G m_{2}}{2 r_{1}r_{12}^{3}} \right. \right. \nn\\
  & \qquad\qquad\qquad\qquad \left.\left. - 
      \frac{3 G m_{2} (n_{12}n_{1})}{4 r_{1}^{2}  r_{12}^{2}} - 
      \frac{G m_{2}}{2 r_{1}^{3} r_{12}} + 
      \frac{G m_{2}}{ 2 r_{2} r_{12}^{3}} \right) - 
    \frac{3 (n_{1}S_{1}v_{1})^{2}}{2 r_{1}^{3}} \right. \nn\\
  & \qquad\qquad \left. + 
    (n_{12}S_{1}S_{1}n_{12}) \left( \frac{15 G m_{2} (n_{12}n_{1})}{4 r_{12}^{4}} - 
      \frac{15 G m_{2} (n_{12}n_{2})}{4 r_{12}^{4}} - 
      \frac{9 G m_{2}}{2 r_{1} r_{12}^{3}} - 
      \frac{3 G m_{2}}{2 r_{2} r_{12}^{3}} \right) \right. \nn\\
  & \qquad\qquad \left. + (n_{1}S_{1}S_{1}n_{1}) 
    \left( -\frac{15 (n_{1}v_{1})^{2}}{4 r_{1}^{3}} + 
      \frac{3 (v_{1}v_{1})}{r_{1}^{3}} + 
      \frac{3 G m_{2} (n_{12}n_{1})}{4 r_{1}^{2 r_{12}^{2}}} + 
      \frac{3 G m_{2}}{2 r_{1}^{3}r_{12}} \right) \right. \nn\\
  & \qquad\qquad \left. + (n_{12}S_{1}S_{1}n_{1}) 
    \left( - \frac{3 G m_{2}}{2 r_{12}^{4}} + 
      \frac{3 G m_{2} }{2 r_{1}^{2} r_{12}^{2}} \right) + 
    \frac{(v_{1}S_{1}S_{1}v_{1})}{r_{1}^{3}} + 
    \frac{3 G m_{2} (n_{12}S_{1}S_{1}n_{2})}{2 r_{12}^{4}} \right] \nn\\
  & + \frac{G}{c^{6}} \left[ (n_{1}S_{1}S_{2}n_{12}) 
    \left( -\frac{3G}{2r_{12}^{4}} - 
      \frac{2G}{r_{1}^{2}r_{12}^{2}}\right) + 
    (n_{12}S_{1}S_{2}n_{12}) \left( \frac{15G (n_{12}n_{1})}{2r_{12}^{4}} - 
      \frac{6G}{r_{1}r_{12}^{3}}\right) \right. \nn\\
  & \qquad\qquad \left. + 
    (S_{1}S_{2}) \left( \frac{-3G (n_{12}n_{1})}{2r_{12}^{4}} + 
      \frac{G}{r_{1}r_{12}^{3}} \right) - 
    \frac{3G(n_{12}S_{1}S_{2}n_{1})}{2r_{12}^{4}} \right] + \exch + \calO(8) \,.
\end{align}

For the calculation of $V_{i}^{\rm SS}$ at its leading order 2PN, we
 proceed similarly as for $V$, but keeping only the first term in the
expansion~\eqref{eq:Vexpand}. The calculation is simpler, with no indirect SS
contributions. We get
\be
V_{i}^{\rm SS} = \frac{G \kappa_{1}}{2m_{1}c^{4}r_{1}^{3}}
\left( 3 (n_{1}S_{1}S_{1}n_{1}) - (S_{1}S_{1}) \right) v_{1}^{i} - 
\frac{G \kappa_{1}}{2m_{1}c^{4}} \frac{4\pi}{3} (S_{1}S_{1}) \delta_{1} + 
\exch + \calO(6) \,,
\ee
where we have included for completeness a distributional term completely analogous to
the one discussed above for the potential $V^{\rm SS}$, but that will not
contribute in the rest of our calculations.

The computation of $\hat{X}^{\rm SS}$ is different, as it involves non-compact
support terms. From Eqs.~\eqref{eq:defpotentials}, we see that the only 1PN
SS contribution in $\hat{X}$ is
\be
\hat{X}^{\rm SS}  = \Box_{\mathcal{R}}^{-1}
\left[ - 2\partial_{i} V_{j}^{\rm SO}\partial_{j} V_{i}^{\rm SO}\right] + 
\calO(4) \,,
\ee
with the leading order SO part of $V_{i}$ given by
\be
V_{i}^{\rm SO} = \frac{G}{2c} S_{1}^{ij}\partial_{j}\frac{1}{r_{1}} + 
\exch + \calO(3) \,.
\ee
We are working at leading order here, so that we keep only the first term in the
expanded inverse d'Alembertian operator, which is just an inverse Laplacian.
For the cross term, with derivatives of both $1/r_{1}$ and $1/r_{2}$, we
use the function $g = \ln\left[ r_{1} + r_{2} + r_{12} \right]$ which
satisfies $\Delta g = 1/(r_{1}r_{2})$ (including the distributional part of the
derivatives). With the notations $\partial^{1}_{i} = \partial/\partial
y_{1}^{i}$, $\partial^{2}_{i} = \partial/\partial y_{2}^{i}$, we
can write\footnote{Including properly the regularization $\mathrm{FP}_{B=0}$ would
  yield an additional constant contribution~\cite{BF00eom}, which would vanish
  after applying the derivatives.}
\be
\Delta^{-1} \left[ \partial_{ij}\left( \frac{1}{r_{1}} \right) 
  \partial_{kl}\left( \frac{1}{r_{2}} \right) \right] = 
\partial^{1}_{ij}\partial^{2}_{kl} 
\left[ \Delta^{-1} \left( \frac{1}{r_{1}r_{2}} \right) \right] = 
\partial^{1}_{ij}\partial^{2}_{kl} g \,.
\ee
For the ``self'' terms, we can ``factorize'' the derivatives as explained
in~\cite{BFP98}. Since we may ignore contributions of the form
$\Delta^{-1}\left( \hat{n}_{1}^{L}/r_{1}^{p} \delta_{1} \right)$ for $\ell+p$
even, we disregard possible distributional terms generated by space or time
differentiation. After factorizing the derivatives, we transform them into
derivatives with respect to $y^i_{1,2}$ and apply $\Delta^{-1}$
straightforwardly on the argument. The relevant formula is
\begin{align}
  \Delta^{-1} \left[ \partial_{ij} 
    \left( \frac{1}{r_{1}} \right)  
    \partial_{kl} \left( \frac{1}{r_{1}} \right) \right] &= 
  \frac{1}{128} \left[ 3\partial_{ijkl}\left( \ln r_{1} \right) - 
    5\left( \delta_{kl}\partial_{ij} + 
      \delta_{ij}\partial_{kl} \right) 
    \left( \frac{1}{r_{1}^{2}} \right) \right. \nn\\
  &\qquad\qquad \left. + 
    3\left( \delta_{ik}\partial_{jl} + \delta_{il}\partial_{jk} + 
      \delta_{jk}\partial_{il} + \delta_{jk}\partial_{il}  \right) 
    \left( \frac{1}{r_{1}^{2}} \right) \right. \nn\\
  &\qquad\qquad \left. + 
    \left( \delta_{ij}\delta_{kl} + \delta_{ik}\delta_{jl} + 
      \delta_{il}\delta_{jk} \right) 
    \Delta \left( \frac{1}{r_{1}^{2}} \right) \right] \,.
\end{align}
Gathering these contributions, we find the following simple expression for the
leading-order SS part of the potential $\hat{X}$:
\be
\hat{X}^{\rm SS} = \frac{G^{2}}{2c^{2}r_{1}^{4}} 
\left[ \frac{1}{4}(S_{1}S_{1}) - (n_{1}S_{1}S_{1}n_{1}) \right] - 
\frac{G^{2}}{2c^{2}} S_{1}^{ij}S_{2}^{kl} 
\partial^{1}_{jk}\partial^{2}_{il} g + \exch + \calO(4) \,,
\ee
where we keep the derivatives in the second term unexpanded.


\subsection{Results for the evolution equations}\label{subsec:resultseom}

Using the results of the previous section and the NS and SO parts of the
metric potentials that are already known, we are in position to complete the
calculation of the equations of motion and precession~\eqref{eq:PFpot}
and~\eqref{eq:Omegapot}. The results for the accelerations $\bm{a}_{1,2}$ and
the precession vectors $\bm{\Omega}_{1,2}$ must pass several tests checking their
validity.

The first one is to make sure of the existence of a set of conserved
quantities, in the absence of reaction reaction at this order, associated with
the Poincar\'e invariance of the problem: a conserved energy $E$, an angular
momentum $\bm{J}$, a linear momentum $\bm{P}$, and a center-of-mass integral
$\bm{G}$. We were actually able to construct all those quantities explicitly
by guess work. The higher-order terms in the precession equations intervene
only in the conservation of the angular momentum, whereas the higher-order
terms in the equations of motion intervene in all other conservation
relations. We shall exhibit below the expression of the conserved energy, which
will be later used to control the phase evolution of the binary in the case of
circular orbits through the balance equation as explained in
Section~\ref{subsec:resultflux}.

Another test consists in checking the Lorentz invariance of the dynamics,
which must be manifest since the harmonic gauge choice is Lorentz-preserving. We
use the same method as in Refs.~\cite{FBB06,MBFB13}, to which we refer the reader
for more details, and find that our results pass this second test.

As the 3PN SS dynamics has been already investigated in both the
EFT~\cite{Porto06,PR06,PR08a,PR08b,Levi08,Levi12} and the
ADM~\cite{SHS07a,SHS07b,SHS08,HSS10,Steinhoff11} approaches, we must be able
to recover their results in our scheme. The equivalence between the ADM and
EFT description has been shown to hold in Refs.~\cite{HSS10,HSS12,LS14a}, so
that we will only compare our results to the ADM ones, in keeping with our
previous works. We present this comparison, and the resulting transformation
from harmonic to ADM variables, in Appendix~\ref{app:adm}. The agreement
with the ADM results also validates the test-mass limit of ours.

Because the expressions produced are rather lengthy, we will give directly their
reduced version in the center-of-mass (CM) frame. As in our
previous works, this frame is defined as the one where
the center-of-mass integral $G_{i}$ (which is such that $\ud
\bm{G} /\ud t = \bm{P}$ and hence $\ud^{2} \bm{G} /\ud t^{2} = 0$) vanishes. We define
$\bm{x} = r\bm{n} = \bm{y}_{1} - \bm{y}_{2}$ the separation vector of the
binary, $\bm{v} = \ud \bm{x} / \ud t$ the relative velocity, $m=m_{1}+m_{2}$
the total mass, $\nu = m_{1}m_{2}/m^{2}$ the symmetric mass ration and $\delta =
(m_{1}-m_{2})/m$ the mass difference. We also use, for convenience, the
same spin variables as in the previous works~\cite{FBB06,BBF06}, namely
\be
\bm{S} = \bm{S}_{1} + \bm{S}_{2} \,, \quad \bm{\Sigma} = 
m\left( \frac{\bm{S}_{2}}{m_{2}} - \frac{\bm{S}_{1}}{m_{1}} \right) \,.
\ee
The vectors $\bm{S}_{1}$ and $\bm{S}_{2}$ are the conserved-norm vectors
constructed in Section~\ref{subsec:defspinvector}. Additionally, we will use the
notation $\kappa_{+} = \kappa_{1}+\kappa_{2}$ and $\kappa_{-} =
\kappa_{1}-\kappa_{2}$.

The positions of the two bodies the new frame read
\be
\bm{y}_{1} = \frac{m_{2}}{m} \bm{x} + 
\frac{1}{c^{2}} \bm{z} \,, \quad \bm{y}_{2} = 
-\frac{m_{1}}{m} \bm{x} + \frac{1}{c^{2}} \bm{z} \,,
\ee
with $\bm{z}$ being a vector related to the center-of-mass integral $\bm{G}$. In
general, when working at the $n$PN order, only the $(n-1)$PN
expression of $\bm{G}$ (or $\bm{z}$) is required. This can be checked explicitly
from the Newtonian expressions in the general frame of the quantities of
interest, as explained for instance in Ref.~\cite{BMFB13}. Thus, we would only need
the SS 2PN expression of $\bm{G}$ in principle, but it turns out that there is no
such contribution in $\bm{G}$. We can therefore translate our results to the
CM frame using simply the same rules as in previous works: namely,
we need the NS 1PN and the SO 1.5PN terms in $\bm{z}$, as
given in the Section~3 of Ref.~\cite{BMFB13}.

For the SS contributions to the conserved energy, we find
\be
E_{\rm SS} = \frac{G\nu}{c^{4}r^{3}} \left[ \frac{1}{4} e_{4}^{0} + 
  \frac{1}{c^{2}} \left( \frac{1}{8}e_{6}^{0} + 
    \frac{1}{4}\frac{G m}{r} e_{6}^{1} \right) \right] \,,
\ee
with
\begin{align}
  e_{4}^{0} &= S^{2} \left(-2 \kappa_{+}-4\right) + 
  (S\Sigma) \left(-2 \delta  \kappa_{+}-4 \delta + 2 \kappa_{-}\right) + 
  \Sigma^{2} \left( \left(\delta  \kappa_{-}-\kappa_{+}\right) + 
    \nu \left(2 \kappa_{+}+4\right) \right)  \nn \\
  &  + (nS)^2 \left(6 \kappa_{+}+12\right) + 
  (n\Sigma)^2 \left( \left(3 \kappa_{+}-3 \delta  \kappa_{-}\right) + 
    \nu \left(-6 \kappa_{+}-12\right) \right)  \nn \\
  &  + (nS) (n\Sigma) \left(6 \delta  \kappa_{+} + 
    12 \delta -6 \kappa_{-}\right) \,, \nn \\
  e_{6}^{0} &= S^{2} \left[(nv)^2 \left( \left(6 \delta  \kappa_{-}-
        6 \kappa_{+}+24\right)  + \nu \left(6 \kappa_{+}+12\right) \right) 
  \right. \nn \\
  & \left. \qquad\qquad  + v^{2} \left( \left(-2 \delta  \kappa_{-} +
        8 \kappa_{+}-28\right)  + 
      \nu \left(2 \kappa_{+}+4\right) \right) \right] \nn \\
  &  + (S\Sigma) \left[(nv)^2 \left( \left(-12 \delta  \kappa_{+} + 
        48 \delta + 12 \kappa_{-}\right)  + \nu \left(6 \delta  \kappa_{+} +
        12 \delta - 30 \kappa_{-}\right) \right)  \right. \nn \\
  & \left. \qquad\qquad  + v^{2} \left( \left(10 \delta  \kappa_{+}- 
        52 \delta -10 \kappa_{-}\right)  + \nu \left(2 \delta  \kappa_{+} + 
        4 \delta + 6 \kappa_{-}\right) \right) \right]  \nn \\
  & + \Sigma^{2} \left[(nv)^2 \left( \left(6 \delta  \kappa_{-} - 
        6 \kappa_{+} + 24\right)  + 
      \nu \left(-9 \delta  \kappa_{-} + 21 \kappa_{+} - 72\right)  + 
      \nu ^2 \left(-6 \kappa_{+} - 12\right) \right) \right. \nn \\
  & \left. \qquad\qquad  + v^{2} \left( \left(-5 \delta  \kappa_{-} +
        5 \kappa_{+} - 24\right)  + \nu \left(\delta  \kappa_{-} - 
        11 \kappa_{+}+76\right)  + \nu ^2 \left(-2 \kappa_{+}-4\right) \right)
  \right]  \nn \\
  &  + (nS)^2 \left[(nv)^2  \nu \left(-30 \kappa_{+}-60\right) + 
    v^{2} \left( \left(60-18 \kappa_{+}\right) + 
      \nu \left(-6 \kappa_{+}-12\right) \right) \right]  \nn \\
  &  + (nS) (vS)(nv) \left( \left(-18 \delta  \kappa_{-} + 
      18 \kappa_{+} - 84\right)  + 
    \nu \left(12 \kappa_{+}+24\right) \right)  \nn \\
  &  + (vS)^2 \left(6 \delta  \kappa_{-}-6 \kappa_{+}+28\right)  \nn \\
  &  + (n\Sigma)^2 \left[(nv)^2 \left( \nu \left(15 \delta  \kappa_{-} - 
        15 \kappa_{+}\right)  + \nu ^2 \left(30 \kappa_{+}+60\right) \right)
  \right. \nn \\
  & \left. \qquad\qquad  + v^{2} \left( \left(9 \delta  \kappa_{-} -
        9 \kappa_{+} + 48\right)  + \nu \left(3 \delta  \kappa_{-} + 
        15 \kappa_{+} - 156\right)  + \nu ^2 \left(6 \kappa_{+} + 
        12\right) \right) \right]  \nn \\
  &  + (n\Sigma) (v\Sigma)(nv) \left( \left(-18 \delta  \kappa_{-} + 
      18 \kappa_{+} - 72\right)  + 
    \nu \left(12 \delta  \kappa_{-} - 48 \kappa_{+} + 228\right)  + 
    \nu ^2 \left(-12 \kappa_{+}-24\right) \right)  \nn \\
  &  + (v\Sigma)^2 \left( \left(6 \delta  \kappa_{-} - 
      6 \kappa_{+} + 24\right) + \nu \left(-6 \delta  \kappa_{-} + 
      18 \kappa_{+} - 76\right) \right)  \nn \\
  &  + (nS) (n\Sigma) \left[(nv)^2 \nu \left(-30 \delta  \kappa_{+} - 
      60 \delta +30 \kappa_{-}\right)  \right. \nn \\
  & \left. \qquad\qquad  + v^{2} \left( \left(-18 \delta  \kappa_{+} + 
        108 \delta +18 \kappa_{-}\right)  + \nu \left(-6 \delta  \kappa_{+} - 
        12 \delta  + 6 \kappa_{-}\right) \right) \right]  \nn \\
  &  + (n\Sigma) (vS)(nv) \left( \left(18 \delta  \kappa_{+} - 
      72 \delta - 18 \kappa_{-}\right) + \nu \left(6 \delta  \kappa_{+} + 
      12 \delta + 30 \kappa_{-}\right) \right)  \nn \\
  &  + (nS) (v\Sigma)(nv) \left( \left(18 \delta  \kappa_{+}- 
      84 \delta - 18 \kappa_{-}\right) + \nu \left(6 \delta  \kappa_{+} + 
      12 \delta + 30 \kappa_{-}\right) \right)  \nn \\
  &  + (vS) (v\Sigma) \left( \left(-12 \delta  \kappa_{+}+ 52 \delta + 
      12 \kappa_{-}\right)  + \nu \left(-24 \kappa_{-}\right) \right) \,, \nn \\
  e_{6}^{1} &= S^{2} \left(-3 \delta  \kappa_{-} + 5 \kappa_{+} + 8\right) + 
  (S\Sigma) \left( \left(8 \delta  \kappa_{+} + 8 \delta - 8 \kappa_{-}\right) 
    + \nu \left(12 \kappa_{-}\right) \right) \nn \\
  &  + \Sigma^{2} \left( \left(4 \kappa_{+} - 
      4 \delta  \kappa_{-}\right) + \nu \left(3 \delta  \kappa_{-} - 
      11 \kappa_{+} - 10\right) \right) + (nS)^2 \left(9 \delta  \kappa_{-} - 
    15 \kappa_{+} - 36\right) \nn \\
  &  + (n\Sigma)^2 \left( \left(12 \delta  \kappa_{-} - 
      12 \kappa_{+}\right) + \nu \left(-9 \delta  \kappa_{-} + 33 \kappa_{+} +
      30\right) \right) \nn \\
  &  + (nS) (n\Sigma) \left( \left(-24 \delta  \kappa_{+} - 32 \delta + 
      24 \kappa_{-}\right)  + \nu \left(-36 \kappa_{-}\right) \right) \,.
\end{align}
The corresponding expressions for the relative acceleration $\bm{a} =
\bm{a}_{1} - \bm{a}_{2}$ and the precession vectors $\bm{\Omega}_{1,2}$ are
provided in Appendix~\ref{app:resultseom}.

Finally, we further specialize our results to the case of circular,
non-precessing orbits. As discussed in Ref.~\cite{BFH12}, we have in fact
three classes of orbits for the conservative dynamics. The CM expression are
valid for general orbits, for which we make no assumption on the presence of
precession and/or eccentricity. Quasi-circular precessing orbits correspond to
the case where we allow a generic orientation of the spins, but assume that
the separation is constant at the SO level; as soon as SS and
higher-order-in-spin terms are included the radius and orbital frequency
become also variable on an orbital timescale. In Ref.~\cite{BFH12} the
definition of such orbits was investigated by perturbing orbital averaged
quantities. The third and simplest class of orbits is that of the circular
orbits with spins aligned with the orbital angular momentum and where precession
is absent. As working at the next-to-leading order makes their determination
more complicated, we leave the investigation of quasi-circular orbits for
future work and focus on the circular, spin-aligned, non-precessing case.

To present results for circular orbits, we use the same definitions as in
previous works. We introduce a moving basis $(\bm{n},\bm{\lambda},\bm{\ell})$,
with $\bm{n}$ denoting the unit vector along the separation vector, $\bm{x} =
r\bm{n}$, $\bm{\ell} = \bm{n}\times \bm{v}/|\bm{n}\times \bm{v}|$ the normal
to the orbital plane, and $\bm{\lambda}$ completing the triad. When neglecting
both radiation reaction and spin precession and assuming the spins aligned
with $\bm{\ell}$ the expressions for the relative velocity and acceleration
become $\bm{v} = r\omega \bm{\lambda}$ and $\bm{a} = -r\omega^{2}\bm{n}$, with
$\omega$ the orbital frequency defined by $\dot{\bm{n}} = \omega\bm{\lambda}$.
For the projected value of the (aligned or anti-aligned) spins along
$\bm{\ell}$ , we use the notation $S_{\ell} = \bm{S}\cdot\bm{\ell}$. We also
introduce the usual PN parameters $\gamma = G m /rc^{2}$ and $x= (G m
\omega/c^{3})^{2/3}$, both of order 1PN. In the following, we only display
the SS terms, and refer the reader to Sections~9.3 and~11.3
of Ref.~\cite{Bliving} for NS and SO contributions, and to Ref.~\cite{Marsat14} for the
newly computed cubic-in-spin contributions.

First, we relate $r$ to $\omega$ by means of the equations of motion. We obtain
the following SS terms for the PN generalization of Kepler's law:
\begin{align}\label{eq:gammacirc}
  \gamma_{\rm SS} &= \frac{x}{G^{2}m^{4}} \left\{ x^{2} 
    \left[ S_{\ell}^2 \left(-\frac{\kappa_{+}}{2}-1\right)  + 
      S_{\ell} \Sigma_{\ell} \left(-\frac{\delta  \kappa_{+}}{2}-\delta +
        \frac{\kappa_{-}}{2}\right) \right.\right. \nn \\
  & \qquad\qquad\qquad\qquad\qquad\qquad \left.\left. + 
      \Sigma_{\ell}^2 \left( \left(\frac{\delta  \kappa_{-}}{4} -
          \frac{\kappa_{+}}{4}\right)  + \nu \left(\frac{\kappa_{+}}{2} + 
          1\right) \right) \right] \right. \nn \\
  & \qquad\qquad\quad \left. + x^{3} \left[ S_{\ell}^2 \left( 
        \left(-\frac{11 \delta  \kappa_{-}}{12} - \frac{11 \kappa_{+}}{12} +
          \frac{14}{9}\right) + \nu \left(-\frac{\kappa_{+}}{6} - 
          \frac{1}{3}\right) \right) \right.\right. \nn \\
  & \qquad\qquad\qquad\qquad \left.\left. + 
      S_{\ell} \Sigma_{\ell} \left( \left(\frac{5 \delta }{3}\right) + 
        \nu \left(-\frac{\delta  \kappa_{+}}{6} - \frac{\delta }{3} +
          \frac{23 \kappa_{-}}{6}\right) \right) \right.\right. \nn \\
  & \qquad\qquad\qquad\qquad \left.\left. + 
      \Sigma_{\ell}^2 \left(1 + \nu \left(\delta  \kappa_{-} - \kappa_{+} - 
          2\right)  + \nu ^2 \left(\frac{\kappa_{+}}{6} +
          \frac{1}{3}\right) \right) \right] + \calO(8) \right\} \,.
\end{align}
The result for the energy for circular, spin-aligned orbits is then
\begin{align}\label{eq:Ecirc}
  E_{\rm SS} &= -\frac{1}{2}m \nu c^{2} x \frac{1}{G^{2} m^{4}} 
  \left\{ x^{2} \left[ \vph{\frac{\delta}{2}} 
      S_{\ell}^2 \left(-\kappa_{+}-2\right)  + 
      S_{\ell} \Sigma_{\ell} \left(-\delta  \kappa_{+}-
        2 \delta +\kappa_{-}\right) \right.\right. \nn \\
  & \qquad\qquad\qquad\qquad\qquad \left.\left. + 
      \Sigma_{\ell}^2 \left( \left(\frac{\delta  \kappa_{-}}{2}-
          \frac{\kappa_{+}}{2}\right)  + 
        \nu \left(\kappa_{+}+2\right) \right)  \right] \right. \nn \\
  & \qquad\qquad\qquad\qquad \left. + x^{3} \left[ S_{\ell}^2 
      \left( \left(-\frac{5 \delta  \kappa_{-}}{3} -
          \frac{25 \kappa_{+}}{6} + \frac{50}{9}\right) + 
        \nu \left(\frac{5 \kappa_{+}}{6} + 
          \frac{5}{3}\right) \right) \right.\right. \nn \\
  & \qquad\qquad\qquad\qquad\qquad \left.\left. + 
      S_{\ell} \Sigma_{\ell} \left( \left(-\frac{5 \delta  \kappa_{+}}{2} +
          \frac{25 \delta }{3} + \frac{5 \kappa_{-}}{2}\right) + 
        \nu \left(\frac{5 \delta  \kappa_{+}}{6} + \frac{5 \delta }{3} +
          \frac{35 \kappa_{-}}{6}\right) \right) \right.\right. \nn \\
  & \qquad\qquad\qquad\qquad\qquad \left.\left. + \Sigma_{\ell}^2 
      \left( \left(\frac{5 \delta  \kappa_{-}}{4} - \frac{5 \kappa_{+}}{4} +
          5\right)  + \nu \left(\frac{5 \delta  \kappa_{-}}{4} + 
          \frac{5 \kappa_{+}}{4} - 10\right)  \right.\right.\right. \nn \\
  & \qquad\qquad\qquad\qquad\qquad\qquad\qquad\qquad\qquad\qquad\qquad\quad
  \left.\left.\left.  + \nu ^2 \left(-\frac{5 \kappa_{+}}{6}-
          \frac{5}{3}\right) \right) \right] \right\} + \calO(8) \,.
\end{align}
This expression can be shown to be in agreement, in the test-mass limit, with
the energy of a test particle in circular equatorial orbits around a Kerr
black hole~\cite{BPT72}. It is crucial to control the phase evolution through
the balance equation (see Section~\ref{subsec:resultflux}).


\section{Next-to-leading order contributions to the post-Newto\-nian
  gravitational waves energy flux} \label{sec:PNflux}

We now move to the computation of the 3PN spin-spin contribution to the energy
flux radiated by the system. We start by briefly reviewing in
Section~\ref{subsec:mpnformalism} the basic elements of the wave generation
formalism that we need here, before providing in Section~\ref{subsec:moments}
some intermediate results useful in the calculation of the source multipole
moments that are required to this order. The explicit results for the moments
in the CM frame are relegated to Appendix~\ref{app:moments} because of their
length. Our explicit result for the GW flux is presented in
Section~\ref{subsec:resultflux} for general orbits in the center of mass in
the system and then reduced to the case of circular orbits in the
configuration where the spins are aligned with the orbital angular momentum.

\subsection{Formalism}
\label{subsec:mpnformalism}

We perform our calculation in the framework of the multipolar post-Newtonian
approach to gravitational radiation. This formalism has been developed over
many years, see e.g.\ ~\cite{Thorne80,BD86,DI91b,BD92,B98mult,PB02}. Since we
will only use a simplified version of the full formalism, as we are working at
next-to-leading order, we will refer the reader to~\cite{Bliving} for a
review, and give only a brief overview.

The asymptotic waveform is defined from the transverse-tracefree (TT)
projection of the metric perturbation, in a suitable radiative coordinate
system $X^\mu=(c\,T,\mathbf{X})$, as its leading-order term in the $1/R$
expansion when the distance $R=\vert\mathbf{X}\vert$ to the source tends to
infinity (keeping the retarded time $T_R\equiv T-R/c$ fixed). It can be
parametrized using two sets of symmetric and trace-free (STF) \emph{radiative}
multipole moments, $U_L$ of mass type and $V_L$ of current type as
\begin{align}\label{waveform}
  h^\mathrm{TT}_{ij} = \frac{4G}{c^2R}
  \,\mathcal{P}^\mathrm{TT}_{ijkl}(\mathbf{N})
  \sum^{+\infty}_{\ell=2}\frac{N_{L-2}}{c^\ell\ell !} \biggl[
  U_{klL-2}(T_R) - \frac{2\ell}{c(\ell+1)}\,N_{m} \,\varepsilon_{mn(k}
  \,V_{l)nL-2}(T_R) \biggr] + \mathcal{O}\left(\frac{1}{R^2}\right)\,,
\end{align}
where we denote by $L=i_1 ... i_\ell$ a multi-index composed of $\ell$
multipolar spatial indices $i_1$, ..., $i_\ell$ ranging from 1 to 3.
Similarly $L-1=i_1 ... i_{\ell-1}$ and $kL-2=k i_1 ... i_{\ell-2}$; $N_L =
N_{i_1}... N_{i_\ell}$ is the product of $\ell$ spatial vectors $N_i$. The
transverse-traceless (TT) projection operator is denoted
$\mathcal{P}^\mathrm{TT}_{ijkl} =
\mathcal{P}_{ik}\mathcal{P}_{jl}-\frac{1}{2}\mathcal{P}_{ij}\mathcal{P}_{kl}$
where $\mathcal{P}_{ij}=\delta_{ij}-N_iN_j$ is the projector orthogonal to the
unit direction $\mathbf{N}=\mathbf{X}/R$ of the radiative coordinate system.
Like in the rest of this paper, the quantity $\varepsilon_{ijk}$ is the
Levi-Civita anti-symmetric symbol such that $\varepsilon_{123}=1$. The
symmetric-trace-free (STF) projection is indicated using brackets or a hat.
Thus $U_L=\hat{U}_L=U_{\langle L\rangle}$ and $V_L=\hat{V}_L=V_{\langle
  L\rangle}$ for STF moments. We denote time derivatives with a superscript
$(n)$.

In terms of these radiative moments, the energy flux into gravitational waves
then reads
\begin{equation}\label{flux}
  \mathcal{F} = \sum_{\ell = 2}^{+ \infty} \frac{G}{c^{2\ell
      +1}}\,\biggl[ \frac{(\ell+1)(\ell+2)}{(\ell-1) \ell \, \ell!
    (2\ell+1)!!} U_L^{(1)} U_L^{(1)} + \frac{4\ell (\ell+2)}{c^2
    (\ell-1) (\ell+1)!  (2\ell+1)!!} V_L^{(1)} V_L^{(1)}\biggr]\,.
\end{equation}
The $U_L$ and $V_L$ can be expressed as (non-linear) functions of two sets of
intermediate source rooted so-called \emph{canonical} moments $M_L$ and $S_L$
which are themselves related by a gauge transformation to a set of two
so-called \emph{source} multipole moments $I_L$, $J_L$ (plus 4 gauge STF
moments) which parametrize the most general solution to the Einstein equations
outside the source. The differences between $M_L$ and $I_L$ (and similarly
between $J_L$ and $S_L$) arise at the 2.5PN order (see for
instance~\cite{BFIS08}) and, since we are interested in SS effects which
always add at least a factor $1/c^2$, we can safely ignore their differences.
Using the same argument, we only need to consider the terms in the relation
between the radiative moments and the canonical ones up to the order 2PN.
Furthermore, we can also neglect the tail terms, which will only generate SS
contributions at the order 3.5PN, so we finally have the simple relation
\bea
(U_{ij})_{\rm SS}&=(I_{ij}^{(2)})_{\rm SS} +\mathcal{O}\left(7\right) \,, \\
(V_{ij})_{\rm SS}&=(J_{ij}^{(2)})_{\rm SS} +\mathcal{O}\left(7\right) \,, \\
(U_{ijk})_{\rm SS}&=(I_{ijk}^{(3)})_{\rm SS} +\mathcal{O}\left(7\right) \,.
\eea
Noticing additionally that the leading order spin-spin contribution to any of
the $I_L$ or $J_L$ (and their time derivatives) is of the order 2PN (as will
be clear from the expressions in the next section), we can express the
spin-spin flux in terms of the relevant source moments as
\begin{align}
\label{fluxSSassourcemoments}
\mathcal{F}_{\rm SS} =&
\frac{G}{c^5}\left\{\frac{1}{5}{I}^{(3)}_{ij}{I}^{(3)}_{ij}
  +\frac{1}{c^2}\left[ \frac{1}{189}{I}^{(4)}_{ijk}{I}^{(4)}_{ijk}
    +\frac{16}{45}{J}^{(3)}_{ij}{J}^{(3)}_{ij}\right]
  +\frac{1}{c^4}\left[\frac{1}{84}{J}^{(4)}_{ijk}{J}^{(4)}_{ijk}\right]
\right\}_{\rm SS}+ \mathcal{O}\left(7\right) \,,
\end{align}
which requires computing the SS parts of $I_{ij}$ to the order 3PN and of
$J_{ij}$ and $I_{ijk}$ to the order 2PN. We also need the NS parts of
$I_{ij}$ up to the order 1PN and of $J_{ij}$ and $I_{ijk}$ at the Newtonian
order, as well as the SO contributions in $I_{ij}$ and $J_{ij}$ up to
the order 1.5PN and the leading 0.5PN SO contribution to $J_{ijk}$, all of
which are known from previous works. Remember that the spin-orbit
contributions to mass (resp. current) type moments start at 1.5PN (resp.
0.5PN) order, and that time derivatives of non-spinning (resp. spin-orbit)
expressions generate spin-spin contributions with an additional order
2PN (resp. 1.5PN) at least.

The matching procedure at the core of the formalism finally allows us to
express the source moments as closed-form integrals over space
~\cite{B98mult}. Instead of reproducing here the general expressions which can
be found in Eq.~(123) of Ref.~\cite{Bliving}, we directly display below the
terms that contribute to the spin-spin corrections at the required orders.
They read
\begin{subequations}
\begin{align}
\label{Iijexplicitintegral}
(I_{ij})_{\rm SS}&=\mathop{\mathrm{FP}}_{B=0}\,\int
\ud^3\mathbf{x}\,\left(\frac{r}{r_0}\right)^B 
\left\{ \hat{x}_{ij} \left[\ov{\Sigma} +\frac{r^2}{14 c^2}
    \ov{\Sigma}^{(2)}\right]-\frac{20}{21 c^2} \hat{x}_{qij}
  \ov{\Sigma}_q^{(1)} \right\}_{\rm SS}+\calO\left(\frac{1}{c^7}\right)\,,\\
(J_{ij})_{\rm SS}&=\mathop{\mathrm{FP}}_{B=0}\,\int \ud^3\mathbf{x}\,
\left(\frac{r}{r_0}\right)^B \varepsilon_{ab<j_\ell} \hat{x}_{i>a} 
\ov{\Sigma}_{b}^{\rm SS}+\calO\left(\frac{1}{c^5}\right)\,, \\
(I_{ijk})_{\rm SS}&=\mathop{\mathrm{FP}}_{B=0}\,\int \ud^3\mathbf{x}\,
\left(\frac{r}{r_0}\right)^B \hat{x}_{ijk} \ov{\Sigma}^{\rm SS}+
\calO\left(\frac{1}{c^5}\right)\,,
\end{align}
\end{subequations}
where $\mathrm{FP}_{B=0}$ denotes a finite part operation defined by analytic
continuation in the complex plane for the parameter $B$, which deals here with
the infrared divergences at infinity. An arbitrary scale $r_0$ is introduced,
which will play no role in the present calculation and has to disappear from
gauge-invariant results. The basic ``building blocks'' $\Sigma$, $\Sigma_i$
and $\Sigma_{ij}$ entering the integrands are defined as
\begin{equation}
\label{Sigma}
\Sigma \equiv \frac{\tau^{00} +\tau^{ii}}{c^2}\,,\qquad\Sigma_i \equiv
\frac{\tau^{0i}}{c}\,,\qquad\Sigma_{ij} \equiv\tau^{ij}\,,
\end{equation} 
where $\tau^{\mu\nu}$ has been defined in~\eqref{eq:Einsteinharmonic}, and the
overline indicates a post-Newtonian (near-zone) expansion. In identifying the
relevant terms in \eqref{Iijexplicitintegral}, we slightly anticipated on the
results of the next subsection (see Eq.~\eqref{eq:Sigmapot}) and used the fact
that the SS contributions to $\Sigma$, $\Sigma_i$ and $\Sigma_{ij}$ all start
at the 2PN order at least.


\subsection{Computation of the source moments}\label{subsec:moments}

To obtain the relevant SS contributions to the source moments, we first
express the sources $\ov{\Sigma}$, $\ov{\Sigma}_i$ and $\ov{\Sigma}_{ij}$ in
terms of the potentials parametrizing the metric and the matter sources
$\sigma$, $\sigma_i$ and $\sigma_{ij}$ defined in \eqref{eq:defsigma} (the
complete relations can be found, generalized to $d$ dimensions,
in~\cite{BDEI05}). Taking into account the order of the spin corrections in
these quantities, the only terms that yield spin-spin contributions to the
orders we are interested in are
\begin{subequations}
\begin{align}
\label{eq:Sigmapot}
\ov{\Sigma}^{\rm SS} &= \left\{ \biggl[1+\frac{4V}{c^2}\biggr]\sigma 
  - \frac{1}{\pi Gc^2}\,\partial_i V \partial_i V + \frac{2}{\pi Gc^4} 
  \partial_i V_j \partial_j V_i \right\}_{\rm SS}+\calO\left(7\right)\,, \\
\ov{\Sigma}_{i}^{\rm SS} &=\left\{  \biggl[ 1 +\frac{4V}{c^2}\biggr] \sigma_i
  - \frac{1}{\pi Gc^2} \partial_k V \partial_k V_i\right\}_{\rm SS} +
\calO\left(5\right)\,,\\
\ov{\Sigma}_{ij}^{\rm SS} &= \calO\left(3\right) \,.
\end{align}
\end{subequations}
The integrals in Eq.~\eqref{Iijexplicitintegral} can now be performed using
the standard techniques described in~\cite{BF00reg,BI04mult}, handling the UV
divergences of the integral through the Hadamard regularization and the IR
divergences through the finite part operation $\mathrm{FP}_{B=0}$.

We highlight here that the distributional parts of the sources have to be
treated with care. In particular, for the first time, we encountered the
situation where such contributions in the metric itself (more precisely in the
potential $V$), and not just those coming from derivatives applied to the
metric, have to be crucially taken into account.

More specifically, the spin-spin leading order contribution in the potential
$V$ was computed in Eqs.~\eqref{eq:Vleading} and~\eqref{eq:Vdistr} and contains a term
proportional to $\delta_1$ which has to be accounted for when integrating the
$\partial_i V \partial_i V$ term of \eqref{eq:Sigmapot} in
\eqref{Iijexplicitintegral}. In order to illustrate this further, let us focus
on the second and third terms in $\ov{\Sigma}_{\rm SS}$
\be
\label{termesdansSigmaecriture1}
\ov{\Sigma}^{\rm SS}_{V} = \frac{4V}{c^2}\sigma - 
\frac{1}{\pi Gc^2}\,\partial_i V \partial_i V \,,
\ee
which we can rewrite using the identity $2 \partial_i A \partial_i
B=\Delta(AB)-A\Delta B -B \Delta A$, and the fact that $\Delta V = - 4\pi G
\sigma$ at leading order, as
\be
\ov{\Sigma}^{\rm SS}_{V} = -\frac{1}{2}\frac{1}{\pi G c^2} \Delta [V^2] \,.
\ee
By reinjecting this second form into \eqref{Iijexplicitintegral}, integrating
by parts, using $\Delta \hat{x}_{ij}=0$ and treating the surface terms as
explained in the Section~IV~D of~\cite{BI04mult}, we readily see that
$\ov{\Sigma}^{\rm SS}_{V}$ actually gives a vanishing contribution to
$I_{ij}$. If on the other hand one uses~\eqref{termesdansSigmaecriture1}
without including the distributional part of $V$, one obtains an incorrect
non-zero result.

Our explicit results for the SS contributions to the source moments reduced to
the center of mass are presented in Appendix~\ref{app:moments}.


\subsection{Gravitational waves energy flux}\label{subsec:resultflux}

Using equation~\eqref{fluxSSassourcemoments}, our results for the source
moments and the equations of motion and precession obtained in
Section~\ref{subsec:resultseom} to compute time derivatives, we can finally
compute explicitly the gravitational wave flux. We will give the result
already reduced in the center-of-mass frame, and we use the same notations as
already introduced in Section~\ref{subsec:resultseom}. We obtain
\be
\calF_{\rm SS} = \frac{G^{3}m^{2}\nu^{2}}{5 c^{9}r^{6}} 
\left[\frac{1}{3}f_{4}^{0} + \frac{1}{21c^{2}} \left( f_{6}^{0} + 
    \frac{G m}{r} f_{6}^{1} + 
    \frac{G^{2} m^{2}}{r^{2}} f_{6}^{2}\right) \right] \,, 
\ee
with
\begin{align}
  f_{4}^{0} &= S^{2} \left[(nv)^2 \left(-312 \kappa_{+}-624\right) + 
    v^{2} \left(288 \kappa_{+}+576\right) \right]  \nn \\
  &  + (S\Sigma) \left[(nv)^2 \left(-312 \delta  \kappa_{+}-
      624 \delta +312 \kappa_{-}\right) + v^{2} \left(288 \delta  \kappa_{+}+
      576 \delta -288 \kappa_{-}\right) \right]  \nn \\
  &  + \Sigma^{2} \left[(nv)^2 \left( \left(156 \delta  \kappa_{-}-
        156 \kappa_{+}+18\right)  + \nu \left(312 \kappa_{+}+
        624\right) \right)  \right. \nn \\
  & \left. \qquad\qquad  + v^{2} \left( \left(-144 \delta  \kappa_{-}+
        144 \kappa_{+}+6\right)  + 
      \nu \left(-288 \kappa_{+}-576\right) \right) \right]  \nn \\
  &  + (nS)^2 \left[(nv)^2 \left(1632 \kappa_{+}+3264\right) + 
    v^{2} \left(-1008 \kappa_{+}-2016\right) \right]  \nn \\
  &  + (nS) (vS)(nv) \left(-696 \kappa_{+}-1392\right) + 
  (vS)^2 \left(144 \kappa_{+}+288\right)  \nn \\
  &  + (n\Sigma)^2 \left[(nv)^2 \left( \left(-816 \delta  \kappa_{-}+
        816 \kappa_{+}+18\right)  + 
      \nu \left(-1632 \kappa_{+}-3264\right) \right)  \right. \nn \\
  & \left. \qquad\qquad  + v^{2} \left( \left(504 \delta  \kappa_{-}-
        504 \kappa_{+}\right)  + 
      \nu \left(1008 \kappa_{+}+2016\right) \right) \right]  \nn \\
  &  + (n\Sigma) (v\Sigma)(nv) \left( \left(348 \delta  \kappa_{-}-
      348 \kappa_{+}-12\right)  + 
    \nu \left(696 \kappa_{+}+1392\right) \right)  \nn \\
  &  + (v\Sigma)^2 \left( \left(-72 \delta  \kappa_{-}+
      72 \kappa_{+}+2\right)  + 
    \nu \left(-144 \kappa_{+}-288\right) \right)  \nn \\
  &  + (nS) (n\Sigma) \left[(nv)^2 \left(1632 \delta  \kappa_{+}+
      3264 \delta -1632 \kappa_{-}\right)  + 
    v^{2} \left(-1008 \delta  \kappa_{+}-2016 \delta +
      1008 \kappa_{-}\right) \right]  \nn \\
  &  + (n\Sigma) (vS)(nv) \left(-348 \delta  \kappa_{+}-
    696 \delta +348 \kappa_{-}\right) + 
  (nS) (v\Sigma)(nv) \left(-348 \delta  \kappa_{+}-696 \delta +
    348 \kappa_{-}\right)  \nn \\
  &  + (vS) (v\Sigma) \left(144 \delta  \kappa_{+}+
    288 \delta -144 \kappa_{-}\right) \,, \nn \\
  f_{6}^{0} &= S^{2} \left[(nv)^4 \left( \left(2274 \delta  \kappa_{-}+
        12918 \kappa_{+}+35436\right)  + 
      \nu \left(-14112 \kappa_{+}-28224\right) \right)  \right. \nn \\
  & \left. \qquad\qquad  + (nv)^2 v^{2} 
    \left( \left(-2592 \delta  \kappa_{-}-17544 \kappa_{+}-51984\right) + 
      \nu \left(17928 \kappa_{+}+35856\right) \right)  \right. \nn \\
  & \left. \qquad\qquad  + v^{4} 
    \left( \left(366 \delta  \kappa_{-}+5034 \kappa_{+}+18276\right) + 
      \nu \left(-4584 \kappa_{+}-9168\right) \right) \right]  \nn \\
  &  + (S\Sigma) \left[(nv)^4 
    \left( \left(10644 \delta  \kappa_{+}+50652 \delta -
        10644 \kappa_{-}\right) \right. \right. \nn \\
  & \left. \left. \qquad\qquad\qquad\qquad + 
      \nu \left(-14112 \delta  \kappa_{+}-28224 \delta +
        5016 \kappa_{-}\right) \right)  \right. \nn \\
  & \left. \qquad\qquad  + (nv)^2 v^{2} 
    \left( \left(-14952 \delta  \kappa_{+}-69672 \delta +
        14952 \kappa_{-}\right) \right. \right. \nn \\
  & \left. \left. \qquad\qquad\qquad\qquad + 
      \nu \left(17928 \delta  \kappa_{+}+35856 \delta -
        7560 \kappa_{-}\right) \right)  \right. \nn \\
  & \left. \qquad\qquad  + v^{4} \left( \left(4668 \delta  \kappa_{+}+
        20812 \delta -4668 \kappa_{-}\right)  + 
      \nu \left(-4584 \delta  \kappa_{+}-9168 \delta +
        3120 \kappa_{-}\right) \right) \right]  \nn \\
  &  + \Sigma^{2} \left[(nv)^4 
    \left( \left(-5322 \delta  \kappa_{-}+5322 \kappa_{+}+9714\right) + 
      \nu \left(4782 \delta  \kappa_{-}-15426 \kappa_{+}-
        64788\right) \right. \right. \nn \\
  & \left. \left. \qquad\qquad\qquad\qquad + 
      \nu ^2 \left(14112 \kappa_{+}+28224\right) \right)  \right. \nn \\
  & \left. \qquad\qquad  + (nv)^2 v^{2} 
    \left( \left(7476 \delta  \kappa_{-}-7476 \kappa_{+}-
        14286\right) \right. \right. \nn \\
  & \left. \left. \qquad\qquad\qquad\qquad + 
      \nu \left(-6372 \delta  \kappa_{-}+21324 \kappa_{+}+86316\right) + 
      \nu ^2 \left(-17928 \kappa_{+}-35856\right) \right)  \right. \nn \\
  & \left. \qquad\qquad  + 
    v^{4} \left( \left(-2334 \delta  \kappa_{-}+2334 \kappa_{+}+3796\right) + 
      \nu \left(1926 \delta  \kappa_{-}-6594 \kappa_{+}-
        23336\right) \right. \right. \nn \\
  & \left. \left. \qquad\qquad\qquad\qquad + 
      \nu ^2 \left(4584 \kappa_{+}+9168\right) \right) \right]  \nn \\
  &  + (nS)^2 \left[(nv)^4 \left( \left(12930 \delta  \kappa_{-}-
        90570 \kappa_{+}-81570\right)  + 
      \nu \left(71520 \kappa_{+}+143040\right) \right)  \right. \nn \\
  & \left. \qquad\qquad  + (nv)^2 v^{2} 
    \left( \left(-8124 \delta  \kappa_{-}+81636 \kappa_{+}+65220\right) + 
      \nu \left(-62976 \kappa_{+}-125952\right) \right)  \right. \nn \\
  & \left. \qquad\qquad  + v^{4} \left( \left(570 \delta  \kappa_{-}-
        14778 \kappa_{+}-6546\right)  + 
      \nu \left(13632 \kappa_{+}+27264\right) \right) \right]  \nn \\
  &  + (nS) (vS) \left[(nv)^3 \left( \left(-19752 \delta  \kappa_{-}+
        51816 \kappa_{+}+16464\right)  + 
      \nu \left(-29184 \kappa_{+}-58368\right) \right)  \right. \nn \\
  & \left. \qquad\qquad  + (nv) v^{2} 
    \left( \left(9522 \delta  \kappa_{-}-19890 \kappa_{+}-180\right) + 
      \nu \left(4092 \kappa_{+}+8184\right) \right) \right]  \nn \\
  &  + (vS)^2 \left[(nv)^2 \left( 
      \left(6378 \delta  \kappa_{-}-9114 \kappa_{+}+7794\right) + 
      \nu \left(5100 \kappa_{+}+10200\right) \right)  \right. \nn \\
  & \left. \qquad\qquad  + v^{2} 
    \left( \left(-1668 \delta  \kappa_{-}-324 \kappa_{+}-6478\right) + 
      \nu \left(120 \kappa_{+}+240\right) \right) \right]  \nn \\
  &  + (n\Sigma)^2 \left[(nv)^4 \left( \left(51750 \delta  \kappa_{-}-
        51750 \kappa_{+}+18420\right)  \right. \right. \nn \\
  & \left. \left. \qquad\qquad\qquad\qquad + 
      \nu \left(-48690 \delta  \kappa_{-}+152190 \kappa_{+}+60960\right) + 
      \nu ^2 \left(-71520 \kappa_{+}-143040\right) \right)  \right. \nn \\
  & \left. \qquad\qquad  + (nv)^2 v^{2} \left( 
      \left(-44880 \delta  \kappa_{-}+44880 \kappa_{+}-
        8112\right) \right. \right. \nn \\
  & \left. \left. \qquad\qquad\qquad\qquad + 
      \nu \left(39612 \delta  \kappa_{-}-129372 \kappa_{+}-79608\right) + 
      \nu ^2 \left(62976 \kappa_{+}+125952\right) \right)  \right. \nn \\
  & \left. \qquad\qquad  + v^{4} 
    \left( \left(7674 \delta  \kappa_{-}-7674 \kappa_{+}+3090\right)  + 
      \nu \left(-7386 \delta  \kappa_{-}+22734 \kappa_{+}+
        10884\right) \right. \right. \nn \\
  & \left. \left. \qquad\qquad\qquad\qquad + 
      \nu ^2 \left(-13632 \kappa_{+}-27264\right) \right) \right]  \nn \\
  &  + (n\Sigma) (v\Sigma) \left[(nv)^3 \left( 
      \left(-35784 \delta  \kappa_{-}+35784 \kappa_{+}-
        33534\right) \right. \right. \nn \\
  & \left. \left. \qquad\qquad\qquad\qquad + 
      \nu \left(34344 \delta  \kappa_{-}-105912 \kappa_{+}+48858\right) + 
      \nu ^2 \left(29184 \kappa_{+}+58368\right) \right)  \right. \nn \\
  & \left. \qquad\qquad  + (nv) v^{2} \left( 
      \left(14706 \delta  \kappa_{-}-14706 \kappa_{+}+7782\right) + 
      \nu \left(-11568 \delta  \kappa_{-}+40980 \kappa_{+}-
        7794\right) \right. \right. \nn \\
  & \left. \left. \qquad\qquad\qquad\qquad + 
      \nu ^2 \left(-4092 \kappa_{+}-8184\right) \right) \right]  \nn \\
  &  + (v\Sigma)^2 \left[(nv)^2 \left( 
      \left(7746 \delta  \kappa_{-}-7746 \kappa_{+}+14124\right) + 
      \nu \left(-8928 \delta  \kappa_{-}+24420 \kappa_{+}-
        36432\right) \right. \right. \nn \\
  & \left. \left. \qquad\qquad\qquad\qquad + 
      \nu ^2 \left(-5100 \kappa_{+}-10200\right) \right)  \right. \nn \\
  & \left. \qquad\qquad  + v^{2} \left( 
      \left(-672 \delta  \kappa_{-}+672 \kappa_{+}-2242\right) + 
      \nu \left(1608 \delta  \kappa_{-}-2952 \kappa_{+}+
        9788\right) \right. \right. \nn \\
  & \left. \left. \qquad\qquad\qquad\qquad + 
      \nu ^2 \left(-120 \kappa_{+}-240\right) \right) \right]  \nn \\
  &  + (nS) (n\Sigma) \left[(nv)^4 \left( 
      \left(-103500 \delta  \kappa_{+}-70920 \delta +
        103500 \kappa_{-}\right) \right. \right. \nn \\
  & \left. \left. \qquad\qquad\qquad\qquad + 
      \nu \left(71520 \delta  \kappa_{+}+143040 \delta -
        123240 \kappa_{-}\right) \right)  \right. \nn \\
  & \left. \qquad\qquad  + (nv)^2 v^{2} \left( 
      \left(89760 \delta  \kappa_{+}+71808 \delta -
        89760 \kappa_{-}\right) \right. \right. \nn \\
  & \left. \left. \qquad\qquad\qquad\qquad + 
      \nu \left(-62976 \delta  \kappa_{+}-125952 \delta +
        95472 \kappa_{-}\right) \right)  \right. \nn \\
  & \left. \qquad\qquad  + v^{4} \left( 
      \left(-15348 \delta  \kappa_{+}-8664 \delta +15348 \kappa_{-}\right) + 
      \nu \left(13632 \delta  \kappa_{+}+27264 \delta -
        15912 \kappa_{-}\right) \right) \right]  \nn \\
  &  + (n\Sigma) (vS) \left[(nv)^3 
    \left( \left(35784 \delta  \kappa_{+}-15402 \delta -
        35784 \kappa_{-}\right) \right. \right. \nn \\
  & \left. \left. \qquad\qquad\qquad\qquad + \nu 
      \left(-14592 \delta  \kappa_{+}-29184 \delta +
        54096 \kappa_{-}\right) \right)  \right. \nn \\
  & \left. \qquad\qquad  + (nv) v^{2} \left( 
      \left(-14706 \delta  \kappa_{+}+8190 \delta +14706 \kappa_{-}\right) + 
      \nu \left(2046 \delta  \kappa_{+}+4092 \delta -
        21090 \kappa_{-}\right) \right) \right]  \nn \\
  &  + (nS) (v\Sigma) \left[(nv)^3 \left( 
      \left(35784 \delta  \kappa_{+}-240 \delta -
        35784 \kappa_{-}\right)  \right. \right. \nn \\
  & \left. \left. \qquad\qquad\qquad\qquad + 
      \nu \left(-14592 \delta  \kappa_{+}-29184 \delta +
        54096 \kappa_{-}\right) \right)  \right. \nn \\
  & \left. \qquad\qquad  + (nv) v^{2} \left( 
      \left(-14706 \delta  \kappa_{+}-5124 \delta +14706 \kappa_{-}\right) + 
      \nu \left(2046 \delta  \kappa_{+}+4092 \delta -
        21090 \kappa_{-}\right) \right) \right]  \nn \\
  &  + (vS) (v\Sigma) \left[(nv)^2 \left( 
      \left(-15492 \delta  \kappa_{+}+23052 \delta +
        15492 \kappa_{-}\right) \right. \right. \nn \\
  & \left. \left. \qquad\qquad\qquad\qquad + 
      \nu \left(5100 \delta  \kappa_{+}+10200 \delta -
        30612 \kappa_{-}\right) \right)  \right. \nn \\
  & \left. \qquad\qquad  + v^{2} \left( 
      \left(1344 \delta  \kappa_{+}-8188 \delta -1344 \kappa_{-}\right) + 
      \nu \left(120 \delta  \kappa_{+}+240 \delta +
        6552 \kappa_{-}\right) \right) \right]\, \nn\\
  f_{6}^{1} &= S^{2} \left[(nv)^2 
    \left( \left(-2772 \delta  \kappa_{-}+23844 \kappa_{+}+48872\right) + 
      \nu \left(-1320 \kappa_{+}-2640\right) \right)  \right. \nn \\
  & \left. \qquad\qquad  + v^{2} \left( \left(2572 \delta  \kappa_{-}-
        21028 \kappa_{+}-41832\right)  + 
      \nu \left(720 \kappa_{+}+1440\right) \right) \right]  \nn \\
  &  + (S\Sigma) \left[(nv)^2 \left( \left(26616 \delta  \kappa_{+}+
        55176 \delta -26616 \kappa_{-}\right)  + 
      \nu \left(-1320 \delta  \kappa_{+}-2640 \delta +
        12408 \kappa_{-}\right) \right)  \right. \nn \\
  & \left. \qquad\qquad  + 
    v^{2} \left( \left(-23600 \delta  \kappa_{+}-48872 \delta +
        23600 \kappa_{-}\right)  + 
      \nu \left(720 \delta  \kappa_{+}+1440 \delta -
        11008 \kappa_{-}\right) \right) \right]  \nn \\
  &  + \Sigma^{2} \left[(nv)^2 \left( \left(-13308 \delta  \kappa_{-}+
        13308 \kappa_{+}+1208\right)  + \nu \left(3432 \delta  \kappa_{-}-
        30048 \kappa_{+}-61736\right) \right. \right. \nn \\
  & \left. \left. \qquad\qquad\qquad\qquad + 
      \nu ^2 \left(1320 \kappa_{+}+2640\right) \right)  \right. \nn \\
  & \left. \qquad\qquad  + v^{2} \left( \left(11800 \delta  \kappa_{-}-
        11800 \kappa_{+}-4408\right)  + \nu \left(-2932 \delta  \kappa_{-}+
        26532 \kappa_{+}+56288\right) \right. \right. \nn \\
  & \left. \left. \qquad\qquad\qquad\qquad + 
      \nu ^2 \left(-720 \kappa_{+}-1440\right) \right) \right]  \nn \\
  &  + (nS)^2 \left[(nv)^2 \left( \left(28788 \delta  \kappa_{-}-
        135588 \kappa_{+}-300528\right)  + 
      \nu \left(2028 \kappa_{+}+4056\right) \right)  \right. \nn \\
  & \left. \qquad\qquad  + v^{2} \left( \left(-12380 \delta  \kappa_{-}+
        77736 \kappa_{+}+182752\right)  + 
      \nu \left(-1992 \kappa_{+}-3984\right) \right) \right]  \nn \\
  &  + (nS) (vS)(nv) \left( \left(-20472 \delta  \kappa_{-}+
      64056 \kappa_{+}+123000\right)  + 
    \nu \left(1932 \kappa_{+}+3864\right) \right)  \nn \\
  &  + (vS)^2 \left( \left(4664 \delta  \kappa_{-}-14652 \kappa_{+}-
      27240\right)  + \nu \left(-168 \kappa_{+}-336\right) \right)  \nn \\
  &  + (n\Sigma)^2 \left[(nv)^2 \left( \left(82188 \delta  \kappa_{-}-
        82188 \kappa_{+}-5604\right)  + \nu \left(-29802 \delta  \kappa_{-}+
        194178 \kappa_{+}+264672\right) \right. \right. \nn \\
  & \left. \left. \qquad\qquad\qquad\qquad + 
      \nu ^2 \left(-2028 \kappa_{+}-4056\right) \right)  \right. \nn \\
  & \left. \qquad\qquad  + v^{2} \left( \left(-45058 \delta  \kappa_{-}+
        45058 \kappa_{+}+9700\right)  + \nu \left(13376 \delta  \kappa_{-}-
        103492 \kappa_{+}-185532\right) \right. \right. \nn \\
  & \left. \left. \qquad\qquad\qquad\qquad + 
      \nu ^2 \left(1992 \kappa_{+}+3984\right) \right) \right]  \nn \\
  &  + (n\Sigma) (v\Sigma)(nv) \left( \left(-42264 \delta  \kappa_{-}+
      42264 \kappa_{+}-9808\right)  + 
    \nu \left(19506 \delta  \kappa_{-}-
      104034 \kappa_{+}-69804\right) \right. \nn \\
  & \left. \qquad\qquad\qquad\qquad + 
    \nu ^2 \left(-1932 \kappa_{+}-3864\right) \right)  \nn \\
  &  + (v\Sigma)^2 \left( \left(9658 \delta  \kappa_{-}-
      9658 \kappa_{+}+4784\right)  + 
    \nu \left(-4580 \delta  \kappa_{-}+
      23896 \kappa_{+}+9808\right) \right. \nn \\
  & \left. \qquad\qquad\qquad\qquad + 
    \nu ^2 \left(168 \kappa_{+}+336\right) \right)  \nn \\
  &  + (nS) (n\Sigma) \left[(nv)^2 
    \left( \left(-164376 \delta  \kappa_{+}-282444 \delta +
        164376 \kappa_{-}\right) \right. \right. \nn \\
  & \left. \left. \qquad\qquad\qquad\qquad + 
      \nu \left(2028 \delta  \kappa_{+}+4056 \delta -
        117180 \kappa_{-}\right) \right)  \right. \nn \\
  & \left. \qquad\qquad  + v^{2} \left( 
      \left(90116 \delta  \kappa_{+}+183716 \delta -90116 \kappa_{-}\right) + 
      \nu \left(-1992 \delta  \kappa_{+}-3984 \delta +
        51512 \kappa_{-}\right) \right) \right]  \nn \\
  &  + (n\Sigma) (vS)(nv) \left( \left(42264 \delta  \kappa_{+}+
      46432 \delta -42264 \kappa_{-}\right)  + 
    \nu \left(966 \delta  \kappa_{+}+1932 \delta +
      39978 \kappa_{-}\right) \right)  \nn \\
  &  + (nS) (v\Sigma)(nv) \left( \left(42264 \delta  \kappa_{+}+
      50744 \delta -42264 \kappa_{-}\right)  + 
    \nu \left(966 \delta  \kappa_{+}+1932 \delta +
      39978 \kappa_{-}\right) \right)  \nn \\
  &  + (vS) (v\Sigma) \left( \left(-19316 \delta  \kappa_{+}-
      18480 \delta +19316 \kappa_{-}\right)  + 
    \nu \left(-168 \delta  \kappa_{+}-336 \delta -
      18488 \kappa_{-}\right) \right) \,, \nn\\
  f_{6}^{2} &= S^{2} \left( \left(16 \delta  \kappa_{-}+ 144 \kappa_{+}+368\right)  + 
    \nu \left(-576 \kappa_{+}-1152\right) \right)  \nn\\
	&  + (S\Sigma) \left( \left(128 \delta  \kappa_{+}+
        224 \delta -128 \kappa_{-}\right)  + \nu \left(-576 \delta\kappa_{+}-
        1152 \delta +512 \kappa_{-}\right) \right)  \nn\\
	&  + \Sigma^{2} \left( \left(-64 \delta  \kappa_{-}+64 \kappa_{+}+24\right)  + 
      \nu \left(272 \delta  \kappa_{-}-400 \kappa_{+}-384\right)  + 
      \nu ^2 \left(576 \kappa_{+}+1152\right) \right)  \nn\\
	&  + (nS)^2 \left( \left(-48 \delta  \kappa_{-}-432 \kappa_{+}-936\right)+ 
      \nu \left(1728 \kappa_{+}+3456\right) \right)  \nn\\
	&  + (n\Sigma)^2 \left( \left(192 \delta  \kappa_{-}-192 \kappa_{+}-16\right)  + 
      \nu \left(-816 \delta  \kappa_{-}+1200 \kappa_{+}+928\right) \right. \nn\\
	& \qquad\qquad\qquad\qquad\qquad\qquad\qquad\qquad\qquad \left. + 
      \nu ^2 \left(-1728 \kappa_{+}-3456\right) \right)  \nn\\
	&  + (nS) (n\Sigma) \left( \left(-384 \delta  \kappa_{+}-
        784 \delta +384 \kappa_{-}\right)  + 
      \nu \left(1728 \delta  \kappa_{+}+3456 \delta -1536 \kappa_{-}\right) \right) \,.
\end{align}

After reduction to the case of spin-aligned, circular orbits, using the
notations already introduced in Section~\ref{subsec:resultseom} for the
energy, we obtain
\begin{align}\label{eq:fluxcirc}
\calF_{\rm SS} &= \frac{32 \nu^{2}}{5} \frac{c^{5}x^{5}}{G}
\frac{1}{G^{2}m^{4}} \left\{ x^{2} \left[ \vph{\frac{1}{16}} S_{\ell}^2
    \left(2 \kappa_{+}+4\right) + S_{\ell} \Sigma_{\ell} \left(2 \delta
      \kappa_{+}+4 \delta -2 \kappa_{-}\right) \right.\right. \nn \\
	& \qquad\qquad\qquad\qquad\qquad\quad \left.\left. + \Sigma_{\ell}^2
        \left( \left(-\delta  \kappa_{-}+\kappa_{+}+\frac{1}{16}\right) + \nu
          \left(-2 \kappa_{+}-4\right) \right) \right] \right. \nn \\
	& \qquad\qquad\qquad\qquad \left. + x^{3} \left[ S_{\ell}^2 \left(
          \left(\frac{41 \delta  \kappa_{-}}{16}-\frac{271
              \kappa_{+}}{112}-\frac{5239}{504}\right) + \nu \left(-\frac{43
              \kappa_{+}}{4}-\frac{43}{2}\right) \right) \right.\right. \nn \\
	& \qquad\qquad\qquad\qquad\qquad \left.\left. + S_{\ell} \Sigma_{\ell}
        \left( \left(-\frac{279 \delta  \kappa_{+}}{56}-
            \frac{817 \delta}{56}+\frac{279 \kappa_{-}}{56}\right) 
        \right.\right.\right. \nn \\
	& \qquad\qquad\qquad\qquad\qquad\qquad\qquad\quad \left.\left.\left. + 
          \nu \left(-\frac{43 \delta  \kappa_{+}}{4}-
            \frac{43 \delta }{2}+\frac{\kappa_{-}}{2}\right) \right)  
      \right.\right. \nn \\
	& \qquad\qquad\qquad\qquad\qquad \left.\left. + 
        \Sigma_{\ell}^2 \left( \left(\frac{279 \delta  \kappa_{-}}{112}-
            \frac{279 \kappa_{+}}{112}-\frac{25}{8}\right)  + 
          \nu \left(\frac{45 \delta  \kappa_{-}}{16}+
            \frac{243 \kappa_{+}}{112}+\frac{344}{21}\right) 
        \right.\right.\right. \nn \\
	& \qquad\qquad\qquad\qquad\qquad\qquad\qquad\quad \left.\left.\left. + 
          \nu ^2 \left(\frac{43 \kappa_{+}}{4}+
            \frac{43}{2}\right) \right)  \right] + \calO(7) \right\} \,.
\end{align}
Using this result as well as the expression of the orbital
energy~\eqref{eq:Ecirc}, we can write the balance equation $\calF = - \ud E/\ud
t$ for circular orbits to obtain the phase evolution of the binary. Different
ways of mixing analytical and numerical integration give rise to different
approximants (see for instance~\cite{Buonanno+09} for a comparison of these
different approximants). For simplicity, we will give here only the phasing
formula for the TaylorT2 approximant: we re-expand $\ud \phi = 2\omega\ud t =
2\omega (-\calF/(\ud E/\ud t))$ and integrate term by term to obtain the phase
of the wave $\phi$ (here $\phi$ is the phase of the leading 22 mode, hence the
factor of 2) as a function of $\omega$ or equivalently of $x$. We get for the
SS contributions
\begin{align}
  (\phi)_{\rm SS} &= -\frac{x^{-5/2}}{32\nu}\frac{1}{G^{2}m^{4}} 
  \left\{ x^{2} \left[ \vph{\frac{1}{2}}S_{\ell}^2 
      \left(-25 \kappa_{+}-50\right)  + 
      S_{\ell} \Sigma_{\ell} \left(-25 \delta  \kappa_{+}-50 \delta +
        25 \kappa_{-}\right) \right.\right. \nn \\
  & \left.\left. \qquad\qquad\qquad\qquad + 
      \Sigma_{\ell}^2 \left( \left(\frac{25 \delta  \kappa_{-}}{2}-
          \frac{25 \kappa_{+}}{2}-\frac{5}{16}\right)  + 
        \nu \left(25 \kappa_{+}+50\right) \right) \right] \right. \nn \\
	& \left. \qquad\qquad\quad + x^{3} \left[ S_{\ell}^2 
        \left( \left(\frac{2215 \delta  \kappa_{-}}{48}+
            \frac{15635 \kappa_{+}}{84}-\frac{31075}{126}\right)  + 
          \nu \left(30 \kappa_{+}+60\right) \right)  \right.\right. \nn \\
	& \left.\left. \qquad\qquad\qquad\quad + 
        S_{\ell} \Sigma_{\ell} \left( 
          \left(\frac{47035 \delta  \kappa_{+}}{336}-
            \frac{9775 \delta }{42}-
            \frac{47035 \kappa_{-}}{336}\right)  \right.\right.\right. \nn \\
	& \left.\left.\left. \qquad\qquad\qquad\qquad\qquad\qquad  + 
          \nu \left(30 \delta  \kappa_{+}+60 \delta -
            \frac{2575 \kappa_{-}}{12}\right) \right)  \right.\right. \nn \\
	& \left.\left. \qquad\qquad\qquad\quad  + 
        \Sigma_{\ell}^2 \left( 
          \left(-\frac{47035 \delta  \kappa_{-}}{672}+
            \frac{47035 \kappa_{+}}{672}-\frac{410825}{2688}\right)  
        \right.\right.\right. \nn \\
	& \left.\left.\left. \qquad\qquad\qquad\qquad\qquad  + 
          \nu \left(-\frac{2935 \delta  \kappa_{-}}{48}-
\frac{4415 \kappa_{+}}{56}+\frac{23535}{112}\right) 
\right.\right.\right. \nn \\
	& \left.\left.\left. \qquad\qquad\qquad\qquad\qquad\qquad + 
          \nu ^2 \left(-30 \kappa_{+}-60\right) \right) 
        \vph{\frac{1}{2}} \right] + \calO(7) \right\} \,.
\end{align}
\begin{table*}[h]
\begin{center}
{\small
\begin{tabular}{|r|c|c|c|}
  \hline
  LIGO/Virgo &  $10 M_{\odot} + 1.4 M_{\odot}$ & $10 M_{\odot} + 10 M_{\odot}$ \\ 
  \hline \hline
  Newtonian &  $3558.9$ & $598.8$ \\ \hline
  1PN & $212.4$ & $59.1$ \\ \hline
  1.5PN & $-180.9+114.0 \chi_1+11.7 \chi_2$ & 
  $-51.2+16.0 \chi_1+16.0 \chi_2$ \\ \hline
  2PN & $9.8 - 10.5 \chi_{1}^{2} - 2.9 \chi_{1}\chi_{2}$ & $4.0 - 1.1
  \chi_{1}^{2} - 2.2 \chi_{1}\chi_{2} - 1.1 \chi_{2}^{2} $ \\  \hline
  2.5PN & $-20.0+33.8 \chi_1+2.9 \chi_2$ & $-7.1+5.7 \chi_1+5.7 \chi_2$ \\
  \hline
  3PN & \begin{tabular}[c]{@{}c@{}} $2.3 - 13.2\chi_1 - 1.3\chi_2 $ \\ 
    $ -1.2 \chi_{1}^{2} - 0.2 \chi_{1}\chi_{2}$ \end{tabular} 
  &  \begin{tabular}[c]{@{}c@{}} $2.2-2.6 \chi_1-2.6 \chi_2 $ \\ 
    $ -0.1 \chi_{1}^{2} - 0.2 \chi_{1}\chi_{2} - 0.1
    \chi_{2}^{2}$ \end{tabular} \\ \hline
  3.5PN & \begin{tabular}[c]{@{}c@{}} $-1.8+11.1 \chi_1+0.8 \chi_2 + 
    (\mathrm{SS}) $ \\ 
    $ - 0.7 \chi_{1}^{3} - 0.3 \chi_{1}^{2}\chi_{2}$ \end{tabular} &  
  \begin{tabular}[c]{@{}c@{}} $-0.8+1.7 \chi_1+1.7 \chi_2 + 
    (\mathrm{SS}) $ \\ 
    $- 0.05 \chi_{1}^{3} - 0.2 \chi_{1}^{2}\chi_{2} - 
    0.2 \chi_{1}\chi_{2}^{2} - 0.05 \chi_{2}^{3}$ \end{tabular} \\ \hline
  4PN & $ (\mathrm{NS}) -8.0 \chi_1 - 0.7 \chi_2 + (\mathrm{SS}) $ & 
  $(\mathrm{NS}) -1.5 \chi_1 - 1.5 \chi_2 + (\mathrm{SS}) $\\ \hline
\end{tabular}
}\end{center}
\caption{
Number of cycles associated to the different PN terms in the phasing formula,
between the starting frequency for advanced detectors (10Hz) and a cut-off
chosen to be the Scwarzschild ISCO $x=1/6$. We show the result for typical
black hole/neutron star and black hole/black hole systems. Spin-aligned,
circular orbits are assumed, and we use the dimensionless spins $\chi_{A}$
such that $S_{A\ell} = G m_{A}^{2} \chi_{A}$. We ignore contributions that are
at least quadratic in the spin of the neutron star. We gather all
contributions known to date, the ones still unknown are indicated in
parenthesis.
\label{tab:cycles}}
\end{table*}
The known NS and SO contributions are summarized in Sections~9.3 and~11.3
of~\cite{Bliving}, and additional cubic-in-spin 3.5PN contributions can
be found in~\cite{Marsat14}. We give in table~\ref{tab:cycles} the number of
cycles of the signal resulting from each term in the phasing formula, for the
frequency band of advanced LIGO/Virgo detectors. Notice however that these
results are illustrative, as they are specific to the TaylorT2 approximant and
as these number of cycles give only a rough idea of the relevance of these
terms in actual data analysis applications.

We can check that our result~\eqref{eq:fluxcirc} is in agreement, in limit of a
test particle orbiting a Kerr black hole, with the result of~\cite{TSTS96}
obtained in the framework of black hole perturbation theory. We leave for
future work the comparison of our results with the so far incomplete results
(given only at the level of the multipole moments) of~\cite{PRR10,PRR12}.
Natural extensions of the work presented here include the investigation of
quasi-circular precessing orbits, the computation of the spherical harmonic
decomposition of the waveform (or, equivalently, the full polarizations
$h_{+,\times}$), and the implementation of these results for the factorized
waveforms of the Effective-One-Body formalism with spins (see e.g.\
~\cite{BD99,DIN08,Pan+10,Taracchini+12}).


\section*{Acknowledgements}
We are thankful to Luc Blanchet, Alessandra Buonanno and Jan Steinhoff for
useful discussions and comments. S.~M. was supported by the NASA grant
11-ATP-046, as well as the NASA grant NNX12AN10G at the University of Maryland
College Park. G.~F. is grateful to the APC for granting him access to the
various facilities of the laboratory during the preparation of this
work.


\appendix

\section{Explicit results for the equations of
  evolution} \label{app:resultseom}

We gather in this appendix explicit results for the equations of motion and
precession obtained in Section~\ref{subsec:resultseom} that are too long to
be shown in the main text. We present them already reduced to the
center-of-mass frame.

For the precession vectors $\Omega_{1,2}^{i}$, we have found simpler to keep the
variables $S_{1,2}$ and $\kappa_{1,2}$ instead of $S,\Sigma$ and
$\kappa_{\pm}$. We get
\be
\left(\Omega_{1}^{i}\right)_{\rm SO} = \frac{G }{c^{3}r^{3}} 
\left[ w^{i}_{3,0} + \frac{1}{c^{2}} \left( w^{i}_{5,0}  + \frac{G m}{r}
    w^{i}_{5,1} \right) \right] + \calO(7) \,,
\ee
with
\begin{align}
  w^{i}_{3,0} &= 3 (nS_{2})n^{i} - S_{2}^{i} + 
  3\kappa_{1} (nS_{1}) n^{i} \left( \frac{1-\delta}{2\nu} - 1 \right) \,, \nn \\
  w^{i}_{5,0} &= S_{2}^{i} \left[ (nv)^2 \left(\frac{3 \delta }{4}-
      \frac{3 \nu }{2}+\frac{15}{4}\right) + 
    v^{2} \left(-\frac{3 \delta }{4}-\frac{\nu }{2}-
      \frac{9}{4}\right)  \right] \nn \\
	& + n^{i} \left[ (nS_{1}) (nv)^2 \kappa_{1} 
      \left(-\frac{15 \nu }{2}\right) + 
      (nS_{2}) (nv)^2 \left(-\frac{15 \delta }{4}+
        \frac{15 \nu }{2}-\frac{15}{4}\right)  \right. \nn \\
	& \left. \qquad  + (nS_{1}) v^{2} \left( 
        \left(-\frac{3 \delta }{2}-
          3 \frac{1-\delta}{2 \nu}+\frac{9}{2}\right)  + 
        \kappa_{1} \left(-\frac{3 \delta }{2}-\frac{3 \nu }{2}+
          \frac{9 }{2}\frac{1-\delta}{2 \nu}-3\right) \right)  \right. \nn \\
	& \left. \qquad  + (nS_{2}) v^{2} \left(\frac{9 \delta }{4}+
        \frac{3 \nu }{2}+\frac{15}{4}\right) + 
      (nv) (vS_{2}) \left(\frac{3 \delta }{4}-\frac{3 \nu }{2}-
        \frac{9}{4}\right)   \right. \nn \\
	& \left. \qquad  + (nv) (vS_{1}) \left( \left(\frac{3 \delta }{4}+
          \frac{3 }{2}\frac{1-\delta}{2 \nu}-\frac{9}{4}\right)  + 
        \kappa_{1} \left(\frac{3 \nu }{2}-
          \frac{9 }{2}\frac{1-\delta}{2 \nu}+
          \frac{9}{2}\right) \right) \right] \nn \\
	& + v^{i} \left[ (nS_{1}) (nv) \left( \left(\frac{3 \delta }{2}+
          3 \frac{1-\delta}{2 \nu}-\frac{9}{2}\right)  + 
        \kappa_{1} \left(\frac{3 \nu }{2}-
          \frac{9 }{2}\frac{1-\delta}{2 \nu}+
          \frac{9}{2}\right) \right)  \right. \nn \\
	& \left. \qquad\qquad  + (nS_{2}) (nv) \left(-\frac{3 \delta }{4}-
        \frac{3 \nu }{2}-\frac{9}{4}\right) + 
      (vS_{2}) \left(\frac{\delta }{2}+2\right) \right. \nn \\
	& \left. \qquad\qquad  + (vS_{1}) \left( \left(-\frac{3 \delta }{4}-
          \frac{3 }{2}\frac{1-\delta}{2 \nu}+\frac{9}{4}\right)  + 
        \kappa_{1} \left(3 \frac{1-\delta}{2 \nu}-3\right) 
      \right) \right] \,, \nn \\
	w^{i}_{5,1} &= n^{i} \left[ (nS_{1}) \left( 
        \left(\frac{5 \delta }{4}+\frac{\nu }{2}-\frac{5}{4}\right)  + 
        \kappa_{1} \left(-\frac{9 \delta }{4}-
          12 \frac{1-\delta}{2 \nu}+
          \frac{57}{4}\right) \right)  \right. \nn \\
	& \left. \qquad  + (nS_{2}) \left(-\frac{3 \delta }{4}+
        \frac{\nu }{2}-\frac{39}{4}\right) \right] + 
    S_{2}^{i} \left[ \frac{13}{4}+\frac{\delta}{4}-\frac{\nu}{2} \right] \,.
\end{align}
For the relative acceleration $a^{i} = a_{1}^{i} - a_{2}^{i}$, we obtain
(coming back to the $S,\Sigma$ and $\kappa_{\pm}$ variables)
\be
\left(a^{i}\right)_{\rm SS} = 
\frac{G}{c^{4}r^{4}m} \left[ \frac{1}{4}\alpha^{i}_{4,0} + 
  \frac{1}{c^{2}} \left( \frac{1}{8}\alpha^{i}_{6,0}  + 
    \frac{1}{4}\frac{G m}{r} \alpha^{i}_{6,1} \right) \right] + \calO(8) \,,
\ee
with
\begin{align}
  \alpha^{i}_{4,0} &= S^{i} \Bigl[ (nS) \left(-12 \kappa_{+}-24\right) + 
  (n\Sigma) \left(-6 \delta  \kappa_{+}-12 \delta +
    6 \kappa_{-}\right) \Bigr] \nn \\
  & + \Sigma^{i} \Bigl[ (nS) \left(-6 \delta  \kappa_{+}-12 \delta +
    6 \kappa_{-}\right) + (n\Sigma) \left( \left(6 \delta  \kappa_{-}-
      6 \kappa_{+}\right)  + 
    \nu \left(12 \kappa_{+}+24\right) \right)  \Bigr] \nn \\
	& + n^{i} \Bigl[ S^{2} \left(-6 \kappa_{+}-12\right) + 
    \Sigma^{2} \left( \left(3 \delta  \kappa_{-}-3 \kappa_{+}\right)  + 
      \nu \left(6 \kappa_{+}+12\right) \right) + 
    (nS)^2 \left(30 \kappa_{+}+60\right) \nn \\
	&  + (S\Sigma) \left(-6 \delta  \kappa_{+}-12 \delta +
      6 \kappa_{-}\right) + (n\Sigma)^2 \left( \left(15 \kappa_{+}-
        15 \delta  \kappa_{-}\right)  + 
      \nu \left(-30 \kappa_{+}-60\right) \right)  \nn \\
	&  + (nS) (n\Sigma) \left(30 \delta  \kappa_{+}+60 \delta -
      30 \kappa_{-}\right)  \Bigr] \,, \nn \\
	\alpha^{i}_{6,0} &= S^{i} \Bigl[ (nS) \left((nv)^2 \left( 
        \left(60 \kappa_{+}-60 \delta  \kappa_{-}\right)  + 
        \nu \left(60 \kappa_{+}+120\right) \right)  \right. \nn \\
	& \left. \qquad\qquad  + v^{2} \left( \left(12 \delta  \kappa_{-}-
          36 \kappa_{+}-48\right)  + 
        \nu \left(-72 \kappa_{+}-144\right) \right) \right)  \nn \\
	&  + (n\Sigma) \left((nv)^2 \left( \left(60 \delta  \kappa_{+}-
          120 \delta -60 \kappa_{-}\right)  + 
        \nu \left(30 \delta  \kappa_{+}+60 \delta +
          90 \kappa_{-}\right) \right)  \right. \nn \\
	& \left. \qquad\qquad  + v^{2} \left( \left(24 \kappa_{-}-
          24 \delta  \kappa_{+}\right)  + 
        \nu \left(-36 \delta  \kappa_{+}-72 \delta +
          12 \kappa_{-}\right) \right) \right)  \nn \\
	&  + (vS)(nv) \left( \left(30 \delta  \kappa_{-}-30 \kappa_{+}+
        84\right)  + \nu \left(-12 \kappa_{+}-24\right) \right)  \nn \\
	&  + (v\Sigma)(nv) \left( \left(-30 \delta  \kappa_{+}+
        132 \delta +30 \kappa_{-}\right)  + 
      \nu \left(-6 \delta  \kappa_{+}-12 \delta -
        54 \kappa_{-}\right) \right)  \Bigr] \nn \\
	& + \Sigma^{i} \Bigl[ (nS) (nv)^2 \left( 
      \left(60 \delta  \kappa_{+}-60 \kappa_{-}\right)  + 
      \nu \left(30 \delta  \kappa_{+}+60 \delta +
        90 \kappa_{-}\right) \right)  \nn \\
	&  + (n\Sigma) (nv)^2 \left( \left(-60 \delta  \kappa_{-}+
        60 \kappa_{+}-120\right)  + 
      \nu \left(30 \delta  \kappa_{-}-150 \kappa_{+}+240\right)  + 
      \nu ^2 \left(-60 \kappa_{+}-120\right) \right)  \nn \\
	&  + (nv) (vS) \left( 
      \left(-30 \delta  \kappa_{+}+48 \delta +30 \kappa_{-}\right)  + 
      \nu \left(-6 \delta  \kappa_{+}-12 \delta -
        54 \kappa_{-}\right) \right)  \nn \\
	&  + (nv) (v\Sigma) 
    \left( \left(30 \delta  \kappa_{-}-30 \kappa_{+}+96\right)  + 
      \nu \left(-24 \delta  \kappa_{-}+84 \kappa_{+}-276\right)  + 
      \nu ^2 \left(12 \kappa_{+}+24\right) \right)  \nn \\
	&  + (nS) v^{2} \left( 
      \left(-24 \delta  \kappa_{+}-24 \delta +24 \kappa_{-}\right)  + 
      \nu \left(-36 \delta  \kappa_{+}-72 \delta +
        12 \kappa_{-}\right) \right)  \nn \\
	&  + (n\Sigma) v^{2} \left( 
      \left(24 \delta  \kappa_{-}-24 \kappa_{+}+24\right)  + 
      \nu \left(24 \delta  \kappa_{-}+24 \kappa_{+}\right)  + 
      \nu ^2 \left(72 \kappa_{+}+144\right) \right)  \Bigr] \nn \\
	& + n^{i} \Bigl[ (nS)^2 (nv)^2 \nu 
    \left(-210 \kappa_{+}-420\right)  + (nv)^2 S^{2} \left( 
      \left(30 \delta  \kappa_{-}-30 \kappa_{+}+120\right)  + 
      \nu \left(30 \kappa_{+}+60\right) \right)  \nn \\
	&  + (nS) (n\Sigma) (nv)^2 \nu 
    \left(-210 \delta  \kappa_{+}-420 \delta +210 \kappa_{-} \right) +
    (vS)^2 \left(6 \delta  \kappa_{-}-6 \kappa_{+}+84\right)  \nn \\
	&  + (n\Sigma)^2 (nv)^2 \left(\nu 
      \left(105 \delta  \kappa_{-}-105 \kappa_{+}\right)  + 
      \nu ^2 \left(210 \kappa_{+}+420\right) \right)  \nn \\
	&  + (nv)^2 (S\Sigma) \left( 
      \left(-60 \delta  \kappa_{+}+240 \delta +60 \kappa_{-}\right)  + 
      \nu \left(30 \delta  \kappa_{+}+60 \delta -
        150 \kappa_{-}\right) \right)  \nn \\
	&  + (nv)^2 \Sigma^{2} \left( 
      \left(30 \delta  \kappa_{-}-30 \kappa_{+}+120\right)  + 
      \nu \left(-45 \delta  \kappa_{-}+105 \kappa_{+}-360\right)  + 
      \nu ^2 \left(-30 \kappa_{+}-60\right) \right)  \nn \\
	&  + (nS) (nv) (vS) \left( 
      \left(-30 \delta  \kappa_{-}+30 \kappa_{+}-420\right)  + 
      \nu \left(60 \kappa_{+}+120\right) \right)  \nn \\
	&  + (n\Sigma) (nv) (vS) \left( \left(30 \delta  \kappa_{+}-
        240 \delta -30 \kappa_{-}\right)  + 
      \nu \left(30 \delta  \kappa_{+}+60 \delta +
        30 \kappa_{-}\right) \right) \nn \\
	&  + (nS) (nv) (v\Sigma) \left( 
      \left(30 \delta  \kappa_{+}-420 \delta -30 \kappa_{-}\right)  + 
      \nu \left(30 \delta  \kappa_{+}+60 \delta +
        30 \kappa_{-}\right) \right)  \nn \\
	&  + (n\Sigma) (nv) (v\Sigma) \left( 
      \left(-30 \delta  \kappa_{-}+30 \kappa_{+}-240\right)  + 
      \nu \left(900-60 \kappa_{+}\right)  + 
      \nu ^2 \left(-60 \kappa_{+}-120\right) \right)  \nn \\
	&  + (vS) (v\Sigma) \left( 
      \left(-12 \delta  \kappa_{+}+132 \delta +12 \kappa_{-}\right)  + 
      \nu \left(-24 \kappa_{-}\right) \right)  \nn \\
	&  + (v\Sigma)^2 \left( 
      \left(6 \delta  \kappa_{-}-6 \kappa_{+}+48\right)  + 
      \nu \left(-6 \delta  \kappa_{-}+18 \kappa_{+}-180\right) \right)  \nn \\
	&  + (nS)^2 v^{2} \left( \left(60 \kappa_{+}+120\right)  + 
      \nu \left(180 \kappa_{+}+360\right) \right)  \nn \\
	&  + S^{2} v^{2} \left( 
      \left(-6 \delta  \kappa_{-}-6 \kappa_{+}-48\right)  + 
      \nu \left(-36 \kappa_{+}-72\right) \right)  \nn \\
	&  + (nS) (n\Sigma) v^{2} \left( 
      \left(60 \delta  \kappa_{+}+120 \delta -60 \kappa_{-}\right)  + 
      \nu \left(180 \delta  \kappa_{+}+360 \delta -
        180 \kappa_{-}\right) \right)  \nn \\
	&  + (n\Sigma)^2 v^{2} \left( 
      \left(30 \kappa_{+}-30 \delta  \kappa_{-}\right)  + 
      \nu \left(-90 \delta  \kappa_{-}+30 \kappa_{+}-120\right)  + 
      \nu ^2 \left(-180 \kappa_{+}-360\right) \right)  \nn \\
	&  + (S\Sigma) v^{2} \left( \left(-72 \delta\right)  + 
      \nu \left(-36 \delta  \kappa_{+}-72 \delta +
        60 \kappa_{-}\right) \right)  \nn \\
	&  + \Sigma^{2}  v^{2} 
    \left(-24 + \nu \left(24 \delta  \kappa_{-}-24 \kappa_{+}+96\right)  + 
      \nu ^2 \left(36 \kappa_{+}+72\right) \right) \Bigr] \nn \\
	& + v^{i} \Bigl[ (nS)^2 (nv) \left( 
      \left(-240 \kappa_{+}\right)  + 
      \nu \left(120 \kappa_{+}+240\right) \right)  \nn \\
	&  + (nv) S^{2} \left( 
      \left(-12 \delta  \kappa_{-}+60 \kappa_{+}-48\right)  + 
      \nu \left(-24 \kappa_{+}-48\right) \right)  \nn \\
	&  + (nS) (n\Sigma) (nv) \left( 
      \left(-240 \delta  \kappa_{+}+240 \delta +240 \kappa_{-}\right)  + 
      \nu \left(120 \delta  \kappa_{+}+240 \delta -
        120 \kappa_{-}\right) \right)  \nn \\
	&  + (n\Sigma)^2 (nv) \left( 
      \left(120 \delta  \kappa_{-}-120 \kappa_{+}+240\right)  + 
      \nu \left(-60 \delta  \kappa_{-}+300 \kappa_{+}-480\right) \right. \nn \\
	& \qquad\qquad\qquad\qquad\qquad\qquad\qquad\qquad \left. + 
      \nu ^2 \left(-120 \kappa_{+}-240\right) \right)  \nn \\
	&  + (nv) (S\Sigma) \left( 
      \left(72 \delta  \kappa_{+}-144 \delta -72 \kappa_{-}\right)  + 
      \nu \left(-24 \delta  \kappa_{+}-48 \delta +
        72 \kappa_{-}\right) \right)  \nn \\
	&  + (nv) \Sigma^{2} \left( 
      \left(-36 \delta  \kappa_{-}+36 \kappa_{+}-96\right)  + 
      \nu \left(24 \delta  \kappa_{-}-96 \kappa_{+}+240\right)  + 
      \nu ^2 \left(24 \kappa_{+}+48\right) \right)  \nn \\
	&  + (nS) (vS) \left( 
      \left(6 \delta  \kappa_{-}+90 \kappa_{+}+84\right)  + 
      \nu \left(-36 \kappa_{+}-72\right) \right)  \nn \\
	&  + (n\Sigma) (vS) \left( 
      \left(42 \delta  \kappa_{+}-42 \kappa_{-}\right)  + 
      \nu \left(-18 \delta  \kappa_{+}-36 \delta +
        6 \kappa_{-}\right) \right)  \nn \\
	&  + (nS) (v\Sigma) \left( 
      \left(42 \delta  \kappa_{+}+36 \delta -42 \kappa_{-}\right)  + 
      \nu \left(-18 \delta  \kappa_{+}-36 \delta +
        6 \kappa_{-}\right) \right)  \nn \\
	&  + (n\Sigma) (v\Sigma) \left( 
      \left(-42 \delta  \kappa_{-}+42 \kappa_{+}-48\right)  + 
      \nu \left(12 \delta  \kappa_{-}-96 \kappa_{+}+12\right)  + 
      \nu ^2 \left(36 \kappa_{+}+72\right) \right) \Bigr] \,, \nn \\
	\alpha^{i}_{6,1} &= S^{i} \Bigl[ (nS) \left( 
      \left(-24 \delta  \kappa_{-}+72 \kappa_{+}+164\right) + 
      \nu \left(36 \kappa_{+}+72\right) \right)  \nn \\
	&  + (n\Sigma) \left( \left(48 \delta  \kappa_{+}+72 \delta -
        48 \kappa_{-}\right)  + \nu \left(18 \delta  \kappa_{+}+
        36 \delta +30 \kappa_{-}\right) \right) \Bigr] \nn \\
	& + \Sigma^{i} \Bigl[ (nS) \left( \left(48 \delta  \kappa_{+}+
        84 \delta -48 \kappa_{-}\right)  + 
      \nu \left(18 \delta  \kappa_{+}+36 \delta +
        30 \kappa_{-}\right) \right)  \nn \\
	&  + (n\Sigma) \left( \left(48 \kappa_{+}-
        48 \delta  \kappa_{-}\right)  + 
      \nu \left(6 \delta  \kappa_{-}-102 \kappa_{+}-148\right)  + 
      \nu ^2 \left(-36 \kappa_{+}-72\right) \right) \Bigr] \nn \\
	& + n^{i} \Bigl[ (nS)^2 \left( 
      \left(48 \delta  \kappa_{-}-192 \kappa_{+}-420\right)  + 
      \nu \left(-96 \kappa_{+}-192\right) \right)  \nn \\
	&  + S^{2} \left( \left(-8 \delta  \kappa_{-}+40 \kappa_{+}+72\right)  + 
      \nu \left(20 \kappa_{+}+40\right) \right)  \nn \\
	&  + (nS) (n\Sigma) \left( 
      \left(-240 \delta  \kappa_{+}-396 \delta +240 \kappa_{-}\right)  + 
      \nu \left(-96 \delta  \kappa_{+}-192 \delta -
        96 \kappa_{-}\right) \right)  \nn \\
	&  + (n\Sigma)^2 \left( 
      \left(120 \delta  \kappa_{-}-120 \kappa_{+}\right)  + 
      \nu \left(240 \kappa_{+}+372\right)  + 
      \nu ^2 \left(96 \kappa_{+}+192\right) \right)  \nn \\
	&  + (S\Sigma) \left( 
      \left(48 \delta  \kappa_{+}+72 \delta -48 \kappa_{-}\right)  + 
      \nu \left(20 \delta  \kappa_{+}+40 \delta +
        12 \kappa_{-}\right) \right)  \nn \\
	&  + \Sigma^{2} \left( \left(24 \kappa_{+}-
        24 \delta  \kappa_{-}\right)  + 
      \nu \left(-2 \delta  \kappa_{-}-46 \kappa_{+}-72\right)  + 
      \nu ^2 \left(-20 \kappa_{+}-40\right) \right) \Bigr] \,.
\end{align}
%


\section{Explicit results for the source multipole moments}
\label{app:moments}

We list in this appendix explicit results for the newly computed SS
contributions to the source moments, the computation of which is described in
Section~\ref{subsec:moments}. We recall that the brackets indicate the STF
projection.

For the mass quadrupole moment, we obtain
\begin{subequations}
\be
\left(I^{ij}\right)_{\rm SS}=
\frac{1}{m c^4 }\left[ \frac{i^{ij}_{4,0}}{2} + 
\frac{\nu}{84 c^2}\left(i^{ij}_{6,0}+\frac{Gm}{r}i^{ij}_{6,1}\right) \right]
\,,
\ee
with
\begin{align}
	i^{ij}_{4,0} &=-\, S^{<i} S^{j>} \left(\delta  \kappa_-+\kappa_+\right)
+4 \, S^{<i} \Sigma^{j>} \nu \kappa_-
+\, \Sigma^{<i} \Sigma^{j>} \nu(\delta  \kappa_--\kappa_+)\,, \nn\\
  	i^{ij}_{6,0} &=29\, S^{<i} S^{j>}  v^2 \left(\delta  \kappa_- -\kappa_+\right)
  +58\, S^{<i} \Sigma^{j>} v^2 \left((\kappa_- -\delta  \kappa_+ )-
    2\nu\kappa_- \right) \nonumber \\ 
  &+\, \Sigma^{<i} \Sigma^{j>} v^2 \left(29(\delta  \kappa_- -\kappa_+ )+
    \nu(-29\delta  \kappa_- +87 \kappa_+ +140 )\right) 
  +66\, (Sv) S^{<i} v^{j>} \left(\kappa_+-\delta  \kappa_-\right) \nonumber \\ 
  &+66\, (\Sigma v) S^{<i} v^{j>} \left((\delta  \kappa_+-\kappa_-)+
    2\nu\kappa_-\right)
  +66\, (Sv) \Sigma^{<i} v^{j>} \left((\delta  \kappa_+-\kappa_-)+
    2\nu\kappa_-\right) \nonumber \\ 
  &+6\, (\Sigma v) \Sigma^{<i} v^{j>} \left(11(\kappa_+-\delta  \kappa_-)+
    \nu(11 \delta  \kappa_--33 \kappa_+-28)\right)
  +22\, S^2\, v^{<i} v^{j>} \left(\delta  \kappa_--\kappa_+\right) \nonumber \\ 
  &+44\, (S\Sigma) v^{<i} v^{j>} \left((\kappa_--\delta  \kappa_+)-
    2\nu\kappa_-\right) \nonumber \\ 
  &+2\, \Sigma^2\, v^{<i} v^{j>} \left(11(\delta  \kappa_--\kappa_+)+
    \nu(-11 \delta  \kappa_-+33 \kappa_++14)\right)\,, \nn\\
    i^{ij}_{6,1} &=6\, (nS)^2 n^{<i} n^{j>} 
  \left((7 \delta  \kappa_-+18 \kappa_++36)- 40\nu(\kappa_++2)\right)
  \nonumber \\
  &+2\, S^2\, n^{<i} n^{j>} \left((-11 \delta  \kappa_-+8 \kappa_+-48)+
    30\nu(\kappa_++2)\right) \nonumber \\ 
  &+6\, (nS) (n\Sigma) n^{<i} n^{j>} \left((11 \delta  \kappa_++36 \delta -
    11\kappa_-)+4\nu(-10 \delta  \kappa_+-20 \delta +3 \kappa_-)\right)
  \nonumber \\ 
  &+3\, (n\Sigma)^2 n^{<i} n^{j>} \left(11(\kappa_+-\delta  \kappa_-)+
    2\nu(13 \delta  \kappa_--24 \kappa_+-36)+
    80\nu^2(\kappa_++2)\right) \nonumber \\ 
  &+2\, (S\Sigma) n^{<i} n^{j>} \left((19 \delta  \kappa_+-48 \delta -
    19\kappa_-)+2\nu(15 \delta  \kappa_++30 \delta +7 \kappa_-)\right) 
\nonumber \\ 
  &+\, \Sigma^2\, n^{<i} n^{j>} \left(19(\kappa_+-\delta  \kappa_-)-
    2\nu(4\delta  \kappa_-+15 \kappa_++8)-
    60\nu^2(\kappa_++2)\right) \nonumber \\ 
  &+12\, (nS) n^{<i} S^{j>} \left((2 \delta  \kappa_--13 \kappa_+-22)+
    5\nu(\kappa_++2)\right) \nonumber \\ 
  &+2\, (n\Sigma) n^{<i} S^{j>} \left(5(-9 \delta  \kappa_+-2 \delta +
    9\kappa_-)+3\nu(5 \delta  \kappa_++10 \delta -13 \kappa_-)\right) 
  \nonumber \\ 
  &+2\, S^{<i} S^{j>} \left(17 \delta  \kappa_-+109 \kappa_+\right) 
\nonumber \\ 
  &+2\, (nS) n^{<i} \Sigma^{j>} \left((-45 \delta  \kappa_+-122 \delta +
    45\kappa_-)+3\nu(5 \delta  \kappa_++10 \delta -
    13\kappa_-)\right) \nonumber \\ 
  &+6\, (n\Sigma) n^{<i} \Sigma^{j>} \left(15(\delta \kappa_--\kappa_+)+
    \nu(-9 \delta  \kappa_-+39\kappa_++44)-
    10\nu^2(\kappa_++2)\right) \nonumber \\ 
  &+8\, S^{<i} \Sigma^{j>} \left(23(\delta \kappa_+-\kappa_-)-
    17\nu\kappa_-\right) \nonumber \\ 
  &+2\, \Sigma^{<i} \Sigma^{j>} \left(46(\kappa_+-\delta  \kappa_-)+
    \nu(-17 \delta  \kappa_--75 \kappa_++56)\right) \,.
\end{align}
\end{subequations}
The current quadrupole moment reads at leading order  (see
also Ref.~\cite{Marsat14} for leading order expressions at any multipolar order)
\begin{align}
\left(J^{ij}\right)_{\rm SS}=\frac{\nu}{2 c^4 m}& \big(-
2 \kappa_- S^{<i} \epsilon^{j>ab} S_a v_b +
(-3 - \delta \kappa_- + \kappa_+) S^{<i} \epsilon^{j>ab} \Sigma_a v_b\big. 
\nonumber\\
& \big. + (-\delta \kappa_- + \kappa_+) \Sigma^{<i} \epsilon^{j>ab} S_a v_b
+ (-\kappa_- + \delta \kappa_+ + 2 \kappa_- \nu) 
\Sigma^{<i} \epsilon^{j>ab} \Sigma_a v_b\big) \,.
\end{align}
Finally, for the mass octupole moment, we find~\cite{Marsat14}
\begin{align}
\left(I^{ijk}\right)_{\rm SS}=\frac{3\nu r}{2 c^4 m} n^{<i} \left(
-2 \kappa_- S^j S^{k>}
+2(\kappa_+ - \delta \kappa_-) S^j \Sigma^{k>}
+(\delta \kappa_+ - \kappa_- + 2 \kappa_- \nu) \Sigma^j \Sigma^{k>}
\right) \,.
\end{align}


\section{Correspondence between the spin vector and spin tensor
  variables}\label{app:spinvectortensor}

This appendix provides the link between the spin tensor and the conserved-norm
spin vector variables which we use to present our PN results. We recall that
the spin tensor variable $S^{ij}$ is the spatial part, in harmonic
coordinates, of the spin tensor introduced in Section~\ref{subsec:mathpap},
and that the spin vector variable has been defined in
Section~\ref{subsec:defspinvector} as $S^{i} = \tilde{S}^{\undl{i}}$, with
$\tilde{S}_{\mu}$ given by Eq.~\eqref{eq:defscovector} and $\undl{i}$ being a spatial
index referring to the tetrad $e_{\undl{\alpha}}^{\ph{\alpha}\mu}$ constructed in the
same section.

We display below the SS contributions to the expression of the spin vector
in terms of the spin tensor, in the general frame. These contributions
complete those computed at the SO order in Ref.~\cite{BMFB13}, Eqs.~(B.1)
(notice that the spin tensor components there were denoted as
$\tilde{S}^{ij}$ instead of $S^{ij}$). We have
\begin{align}
  \left( S^{i}_{1} \right)_{\rm SS} &= 
  \frac{G}{c^{5}r_{12}^{2}} \left[ 
    2 n_{12}^{a} v_{1}^{b} S_{2}^{aj} S_{1}^{jk} \varepsilon^{ibk}+
    \frac{1}{2} v_{1}^{i} \left(
      n_{12}^{a} S_{2}^{aj} S_{1}^{bi} \varepsilon^{bij}\right)-
    \frac{1}{2} v_{2}^{i} \left(
      n_{12}^{a} S_{2}^{aj} S_{1}^{bi} \varepsilon^{bij}\right) \right. \nn \\ 
  & \left. \qquad\qquad -
      \frac{1}{2} n_{12}^{a} S_{2}^{ia} \left(
        v_{1}^{a} S_{1}^{bi} \varepsilon^{abi}\right)-
      S_{1}^{ab} \varepsilon^{iab} \left(
        n_{12}^{a} v_{1}^{b} S_{2}^{ab}\right)+
      \frac{1}{2} n_{12}^{a} S_{2}^{ia} \left(
        v_{2}^{a} S_{1}^{bi} \varepsilon^{abi}\right) \right. \nn \\ 
	& \left. \qquad\qquad + 
      S_{1}^{ab} \varepsilon^{iab} \left(n_{12}^{a} v_{2}^{b} S_{2}^{ab}\right) 
      \vph{\frac{1}{1}} \right] + \calO(7) \,,
\end{align}
where we have authorized the repetition of indices appearing in scalar
quantities enclosed with parenthesis. At this order, there appear
$S_{1}S_{2}$ terms only and thus no $S_{1}^{2}$, $S_{2}^{2}$ terms.


\section{Equivalence with ADM results for the dynamics} 
\label{app:adm}

In this appendix, we compare our results for the dynamics with those
previously obtained in the ADM~\cite{SHS07a,SHS07b,SHS08,HSS10,Steinhoff11}
and EFT~\cite{Porto06,PR06,PR08a,PR08b,Levi08,Levi12} approaches. As the
equivalence of ADM and EFT results has been already demonstrated in
Refs.~\cite{HSS10,HSS12,LS14a}, we actually restrict ourselves to the
comparison of our findings with the ADM ones, in line with our previous works.

The two results have been obtained in different gauges and the spin variables
differ in their definition. It is thus important to take properly into account
the transformation of the particle positions and spins from one
formalism to the other. In the following, we will denote the ADM variables
with an overbar and resort to the convenient notation
$\ov{\bm{\pi}}_{A}=\ov{\bm{p}}_{A}/m_{A}$. Let us now introduce the contact
transformation $\bm{Y}_{A}(\ov{\bm{x}},\ov{\bm{p}},\ov{\bm{S}})$ and the
rotation vector $\bm{\theta}_{A}(\ov{\bm{x}},\ov{\bm{p}},\ov{\bm{S}})$ such
that the harmonic variables are related to the ADM ones by
\begin{subequations}\label{eq:admtrans}
\begin{align}
\bm{y}_{A} &= \bm{Y}_{A}(\ov{\bm{x}},\ov{\bm{p}},\ov{\bm{S}}) + \calO(7) \,, \\
	\bm{S}_{A} &= \ov{\bm{S}}_{A} + \bm{\theta}_{A}(\ov{\bm{x}},\ov{\bm{p}},
\ov{\bm{S}}) \times \ov{\bm{S}}_{A} +\calO(6) \,.
\end{align}
\end{subequations}
The ADM spin variables and ours have the same Euclidean norm
$\bm{S}_{A}\cdot\bm{S}_{A} = \ov{\bm{S}}_{A}\cdot\ov{\bm{S}}_{A} = s_{A}^{2}$,
which is precisely the conserved norm introduced in Section~\ref{subsec:mathpap}.
Since the first corrections enter as $\bm{\theta}_{A}=\calO(4)$, we see
that the transformation for the spins necessarily takes this form.

Now, if we denote by $\bm{A}_{A}(\ov{\bm{x}},\ov{\bm{p}},\ov{\bm{S}})$ and
$\bm{\Omega}_{A}(\ov{\bm{x}},\ov{\bm{p}},\ov{\bm{S}})$ the function that
converts to ADM variables the harmonic-coordinate acceleration and precession
vector, and by $\ov{\bm{\Omega}}_{A}$ the precession vector of the ADM spins,
such that $\ud\ov{\bm{S}}_{A}/\ud t = \ov{\bm{\Omega}}_{A}\times
\ov{\bm{S}}_{A} $, the two relations to impose for the dynamics to be
equivalent are
\begin{subequations}
\begin{align}
	&\bm{A}_{A} = \left\{ \left\{ \bm{Y}_{A}, H_{\rm ADM} \right\} , 
H_{\rm ADM} \right\} + \calO(7) \,, \\
	& \{\bm{\theta}_{A}, H_{\rm ADM}\} + 
\bm{\theta}_{A}\times\ov{\bm{\Omega}}_{A} = 
\bm{\Omega}_{A}\left( \ov{\bm{x}},\ov{\bm{p}},\ov{\bm{S}} \right) - 
\ov{\bm{\Omega}}_{A} + \calO(6) \,,
\end{align}
\end{subequations}
where $H_{\rm ADM}$ is the ADM Hamiltonian (which can be found for instance in
Section~6.2 of Ref.~\cite{Steinhoff11}) and $\{,\}$ is the usual Poisson brackets
extended to spin variables. Here the term
$\bm{\theta}_{A}\times\ov{\bm{\Omega}}_{A}$ is actually negligible, for
$\bm{\theta}_{A} = \calO(2)$ and $\ov{\bm{\Omega}}_{A} = \calO(4)$.

We find that there are no contributions at leading order in the
transformations~\eqref{eq:admtrans}, i.e.\ $(\bm{Y}_{A})_{\rm SS} = \calO(6)$
and $(\bm{\theta}_{A})_{\rm SO} = \calO(5)$. Using the method of undetermined
coefficients then leads to a unique solution for the higher-order terms in the
transformations. For the rotation vector we obtain
\be
\left( \bm{\theta}_{1} \right)_{\rm SO} = 
\frac{G}{c^{5}\ov{r}_{12}^{2}} \left[ 
  \frac{m_{2}}{m_{1}} \bm{\theta}_{1}^{5,1} + 
  \bm{\theta}_{1}^{5,2} \right] + \calO(7) \,,
\ee
with (adopting the same notations as in the rest of the paper for scalar products)
\begin{align}
	\bm{\theta}_{1}^{5,1} &= -
    \frac{3\kappa_{1}}{2}\ov{\bm{n}}_{12} \left[ 
      (\ov{\pi}_{2}\ov{S}_{1}) + 
      (\ov{n}_{12}\ov{\pi}_{2})(\ov{n}_{12}\ov{S}_{1}) \right] - 
    \frac{3\kappa_{1}}{2}\ov{\bm{\pi}}_{2}(\ov{n}_{12}\ov{S}_{1}) - 
    \frac{1}{2}\ov{\bm{\pi}}_{1}(\ov{n}_{12}\ov{S}_{1})  \,,\nn \\
	\bm{\theta}_{1}^{5,2} &= -
    \frac{1}{2}\ov{\bm{S}}_{2} (\ov{n}_{12}\ov{\pi}_{2}) + 
    \frac{1}{2} \ov{\bm{n}}_{12} \left[ (\ov{\pi}_{2}\ov{S}_{2}) - 
      3(\ov{n}_{12}\ov{\pi}_{2})(\ov{n}_{12}\ov{S}_{2}) \right] + 
    \frac{1}{2} \ov{\bm{\pi}}_{2}(\ov{n}_{12}\ov{S}_{2}) \,.
\end{align}
We recall that SO terms in $\theta$ actually correspond to SS effects in the
dynamics. For the contact transformation, we arrive at the simple expression
\be
\left( \bm{Y}_{1} \right)_{\rm SS} = 
\frac{G m_{2}}{2m_{1}^{2}c^{6}\ov{r}_{12}^{2}} 
\left[ \ov{\bm{S}}_{1}(\ov{n}_{12}\ov{S}_{1}) - 
\ov{\bm{n}}_{12} (\ov{S}_{1}\ov{S}_{1}) \right] + \calO(8) \,.
\ee
The relevant NS and SO contributions to these transformations are given for
instance in Ref.~\cite{MBFB13} and Refs.~\cite{MBFB13,BMFB13}.

The existence of a solution relating our variables to the ADM ones validates
our results, the problem of finding such transformations being largely
over-constrained.


\bibliography{ListeRef}

\end{document}